\documentclass{article}

\usepackage{titlesec}
\usepackage[left=3cm, right=2cm, top=2.5cm, bottom= 2.5cm]{geometry}
\usepackage{setspace} 
\usepackage{amsmath}
\usepackage{graphicx}
\usepackage[colorlinks=true, allcolors=blue]{hyperref}
\usepackage{enumerate}
\usepackage{enumitem}
\usepackage{tabularx}
\usepackage[round]{natbib}
\usepackage{parskip}

\usepackage{xcolor, color, soul}
\sethlcolor{yellow}
\usepackage{amsfonts}
\usepackage{bbm}
\usepackage{float}
\graphicspath{ {./images/}}
\usepackage{authblk}
\usepackage{accents}
\usepackage{graphicx}
\usepackage{booktabs}
\usepackage{hyperref}
\usepackage{etoolbox}
\usepackage{dsfont}

\DeclareMathOperator{\logit}{logit}
\DeclareMathOperator{\E}{\mathbb{E}}

\usepackage{booktabs} 
\usepackage{caption} 
\usepackage{float} 
\usepackage{siunitx}
\usepackage{listings}
\usepackage{xcolor}

\definecolor{rbg}{RGB}{247,247,247}
\definecolor{rcomment}{RGB}{150,90,20}
\definecolor{rnumber}{RGB}{0,0,180}
\definecolor{rstring}{RGB}{0,120,0}
\definecolor{rkeyword}{RGB}{20,60,160}

\lstdefinestyle{Rstyle}{
    language=R,
    backgroundcolor=\color{rbg},
    basicstyle=\ttfamily\small,
    commentstyle=\color{rcomment}\itshape,
    keywordstyle=\color{rkeyword},
    stringstyle=\color{rstring},
    numbers=none,
    frame=none,
    breaklines=true,
    showstringspaces=false,
    columns=fullflexible,
    keepspaces=true,
    xleftmargin=0pt,
    aboveskip=0.5em,
    belowskip=0.5em
}

\title{On the use of G-computation in small randomized controlled trials with many covariates}
\author[1,2]{Muluneh Alene}
\author[1,3]{Kelly Van Lancker}
\author[1]{Stijn Vansteelandt}
\affil[1]{Department of Mathematics, Computer Science and Statistics, Ghent University, Ghent, Belgium}
\affil[2]{Department of Public Health, Debre Markos University, Debre Markos, Ethiopia}
\affil[3]{Department of Mathematics and Data Science, Vrije Universiteit Brussel, Brussels, Belgium}

\begin{document}
	\maketitle

\begin{center}
	\textbf{Abstract}\\
\end{center}

\begin{spacing}{1}
In contemporary randomized controlled trials (RCTs), the number of patients is often small relative to the number of baseline covariates collected. In such settings, maximum likelihood estimators of conditional treatment effects in generalized linear models, along with their standard errors, may exhibit substantial bias. This study examines whether similar bias arises in G-computation estimators of marginal treatment effects, which are known for their robustness to model misspecification when applied with canonical GLMs. We develop theoretical insights, drawing on recent literature on G-computation under proportional asymptotic regimes, in which the number of covariates grows with the sample size. Specifically, we characterize the bias of the standard G-computation estimator and the leave-one-out G-computation estimator under such high-dimensional settings.  Monte Carlo simulations using linear and logistic outcome models are conducted to evaluate practical remedies for G-computation estimators, including covariate selection, cross-fitting, leave-one-out cross-fitting, and small sample corrections to standard errors. Further insights are derived from a re-analysis of the BestAIR trial data. Our findings provide guidance for the application of G-computation in modern RCTs, particularly when addressing challenges posed by limited sample size.

\end{spacing}

\clearpage
\section{Introduction}\label{sec:introduction_ch3}
In the analysis of randomized controlled trials (RCTs), adjusting for baseline covariates is a standard strategy to increase the precision of treatment effect estimates. Traditionally, this adjustment is performed by fitting a generalized linear model (GLM) to estimate conditional treatment effects. However, as noted by \cite{robins1994correcting, freedman2008regression, berk2004regression}, among others, the validity of these model-based conclusions fundamentally relies on the assumption that the statistical model accurately represents the true data-generating distribution. In clinical research, where treatment responses are often influenced by complex interactions and non-linear relationships between patient characteristics, simple parametric models may lack the necessary complexity to capture the true underlying data-generating process.

These concerns regarding model misspecification, combined with a historical lack of clarity regarding the distinction between conditional and marginal estimands, have led to an increasing interest in marginal treatment effects \citep{tsiatis2008covariate, rosenblum2009using, rosenblum2010simple, FDA2023}. Marginal effects are often more directly aligned with the primary scientific and regulatory objectives of a trial. Recent methodological developments have clarified this distinction, leading to the widespread adoption of G-computation estimators based on GLMs with canonical link functions \citep{tsiatis2008covariate, FDA2023}. Its primary appeal in the context of RCTs is robustness: in large samples, these estimators can provide consistent estimates of marginal treatment effects even when the underlying outcome model is misspecified \citep{tsiatis2008covariate}.

However, a separate and increasingly prevalent challenge arises in modern RCTs: the ``small-n, large-p’’ regime \citep{sur2019modern, cherlin2024utilising}. While the robustness of G-computation is well-established asymptotically \citep{tsiatis2008covariate, moore2009covariate}, its performance is not well studied in small samples with a relatively large number of covariates. In contemporary clinical research, the number of baseline covariates being collected can be large relative to the number of patients randomized. Concrete examples are \cite{ibrahim2020multi} ($n=55$, $p=26$), \cite{nzesi2023impact} ($n=64$, $p=38$), \cite{kim2025effect} ($n=28$, $p=17$), and \cite{zhao2017effect} ($n=169$, $p=114$), as well as studies measuring thousands of gene variants in fewer than 100 patients \citep{obi2020effect}. See also \autoref{figure:appe_k_Bingkai_ch3} in Section \ref{appe_data_analysis_results_ch3} for the distribution of the ratio of the number of covariates to the sample size in fifty RCTs across different areas (e.g., cardiovascular, infectious disease and nephrology), as reported by \cite{shao2026benchmarking}. This trend creates significant regulatory uncertainty; for instance, the \cite{FDA2023} guidance explicitly advises that ``sponsors should discuss their proposal with the relevant review division if the number of covariates is large relative to the sample size.'' This cautionary stance stems from the risk that overfitted models may yield unreliable treatment effect estimates and distorted inference.

\begin{figure}[!h]
    \centering
    \includegraphics[width=1\textwidth]{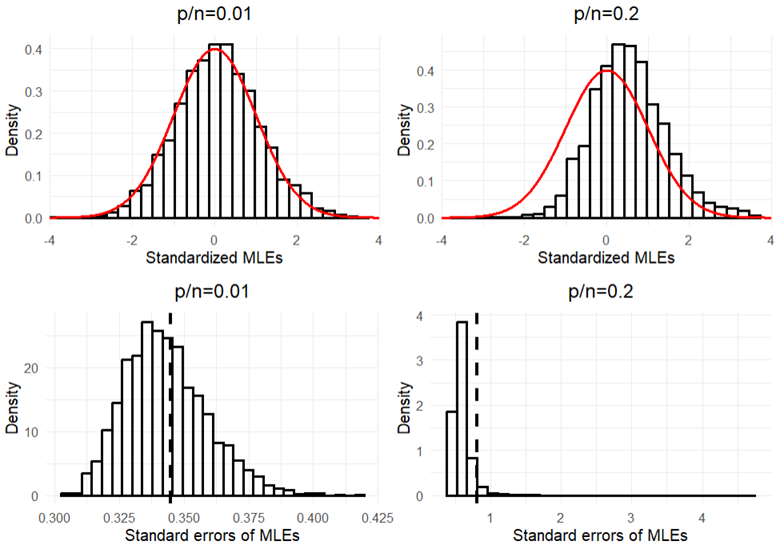}
   \caption{Distribution of the MLEs (standardized) of the treatment coefficient and their model-based standard errors in logistic regression. \textit{Left panels:} Results for $p = 2$ and $n = 200$. \textit{Right panels:} Results for $p = 40$ and $n = 200$. The curve represents the standard normal distribution. Further, the vertical dashed line represents the Monte Carlo standard deviation. Results are based on 3000 simulation replicates. Details of the data-generating mechanisms are provided in Section \ref{sec:simulations_ch3} (Setting 1 with binary outcomes). The standardized MLE estimate is calculated as the difference between the MLE estimate and the treatment coefficient, divided by the standard deviation of the MLE estimates. The density is computed by dividing the number of observations in each bin by the product of the total number of observations and the bin width. The histograms of the standardized MLEs show that the MLEs exhibit increasingly upward bias as the covariate-to-sample size ratio increases. Similarly, the histograms of the model-based standard errors indicate that they increasingly underestimate the Monte Carlo standard deviation of the MLEs as the covariate-to-sample size ratio grows.}
    \label{figure:dist_MLE_bin}
\end{figure}

In these settings, maximum likelihood estimators (MLE) and their standard errors in GLMs are prone to substantial overfitting bias, even with a moderate number of covariates \citep{portnoy1988asymptotic, firth1993bias,sur2019modern, yadlowsky2021sloe, montesinos2022multivariate,lewis2024inference}. This phenomenon is illustrated in \autoref{figure:dist_MLE_bin}, which demonstrates that the MLE of the treatment effect exhibits significant bias when the sample size is 200 and the number of covariates is 40. Beyond bias in the effect estimate itself, overfitting also distorts uncertainty quantification; model-based standard errors, the default in most statistical software, tend to systematically underestimate the true variability of the MLE (see \autoref{figure:dist_MLE_bin}).

\begin{figure}[!h]
    \centering
    \includegraphics[width=0.8\textwidth]{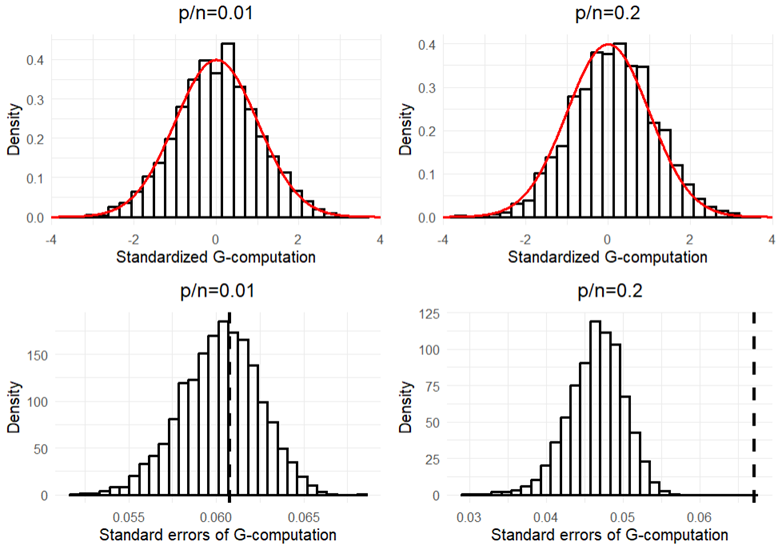}
    \caption{Distribution of the G-computation estimates (standardized) of the ATE and their standard errors with binary outcomes. \textit{Left panels:} Results for $p = 2$ and $n = 200$. \textit{Right panels:} Results for $p = 40$ and $n = 200$. The curve represents the standard normal distribution. Further, the vertical dashed line represents the Monte Carlo standard deviation. Results are based on 3000 simulation replicates. Details of the data-generating mechanisms are provided in Section \ref{sec:simulations_ch3} (Setting 1 with binary outcomes). The standardized estimate is calculated as the difference between the G-computation estimate and the ATE, divided by the standard deviation of the G-computation estimates. The density is computed by dividing the number of observations in each bin by the product of the total number of observations and the bin width. The histograms of the standardized G-computation estimates indicate that G-computation remains unbiased as the covariate-to-sample size ratio increases. However, the histograms of the standard errors show that the standard errors increasingly underestimate the Monte Carlo standard deviation of the G-computation estimates as the covariate-to-sample size ratio grows.}
    \label{figure:dist_G_Comp_bin}
\end{figure}

This raises a critical question: does the overfitting bias inherent in the conditional model propagate to the marginal treatment effect estimators produced via G-computation? While G-computation is designed to be robust to model misspecification in large samples, in small RCTs it may inherit the finite-sample deficiencies of the outcome predictions on which it relies \citep{zhao2026adjusting}. Specifically, the instability of the conditional MLEs may bias marginal point estimates (see, for example, Figure 1 in \cite{zhao2026adjusting}), though the impact on bias appears much less severe (see \autoref{figure:dist_G_Comp_bin}). More importantly, as demonstrated in \autoref{figure:dist_G_Comp_bin}, standard (large-sample) variance estimators may systematically underestimate the true variability in these high-dimensional settings. This can lead to severely distorted inference and inflated type I error rates.

We provide accessible insights into methods, their connections, and reasons for their performance through independent simulation studies. Our findings aim to guide the application of G-computation estimators in modern RCTs, particularly in settings where covariate adjustment is performed with small sample sizes.

\section{Background}\label{sec:background_ch3}
\subsection{Relation to the prior literature}\label{subsec:literature_ch3}

The challenge of covariate adjustment in the ``small-$n$, large-$p$'' regime has sparked a surge of recent methodological developments. These can be broadly categorized into approaches addressing point estimate overfitting and those addressing the systematic underestimation of variability.

Motivated by the need to adjust for many covariates in RCTs, a growing body of literature combines outcome regression with regularization. Under a finite-population randomization model, \cite{bloniarz2016Lasso} established the formal properties of a Lasso-based G-computation estimator for the ATE. They demonstrated that fitting separate $\ell_1$-penalized linear regressions within each treatment arm is more efficient than the simple difference-of-means, provided an ``ultra-sparsity'' assumption holds (namely, that the number of non-zero coefficients grows as $o(\sqrt{n}/\log p)$). This framework was generalized under a superpopulation framework by \cite{wager2016high}, who relaxed these structural constraints to allow the number of non-zero coefficients to grow as $o(n/\log p)$. They showed that efficient ATE estimation is achievable via any prediction-risk consistent adjustment method (e.g., Elastic Net or subset selection) and proposed a cross-fitting procedure to ensure valid inference with non-Gaussian or heteroscedastic outcomes. 

Expanding on these foundations, a variety of recent works have explored alternative penalization schemes and data-adaptive selection procedures to optimize the bias-variance tradeoff in high dimensions \citep{van2024automated, liu2026coadvise, bannick2025general}. For instance, \cite{liu2026coadvise} provides a comprehensive evaluation of variable selection methods, with a focus on AIPW estimators based on both linear and non-linear models. They showed that the AIPW estimator obtained by fitting treatment specific conditional outcome models using OLS, after use of any method of variable selection, is at least as efficient as the unadjusted estimator under the condition that the number of non-zero coefficients in each outcome model grows as $o\left(\sqrt{n}/\log\{\max(p,n)\}\right)$; they provide similar results for non-linear GLMs, albeit under additional conditions on the degree of possible model misspecification.

Beyond variable selection, penalty-based bias reduction, in particular the Firth correction \citep{firth1993bias}, has long been a staple for stabilizing conditional effects. Its application to G-computation, however, is nuanced; while it prevents the non-existence of the MLE in sparse or high-dimensional settings, it may introduce specific prediction biases that propagate directly to the marginal point estimate \citep{zhang2025bias}. Similar challenges arise with Bayesian logistic regression using weakly informative priors (e.g., Cauchy priors), which prevent extreme coefficient estimates but can shrink the resulting ATE estimator toward the null \citep{gelman2008weakly}. Because G-computation is a pure ``plug-in'' estimator, it is highly sensitive to such regularization; to mitigate this, an AIPW framework is often preferred due to its inherent bias-correction properties. In the context of an RCT where treatment probabilities are known, the AIPW estimator's augmentation term acts as a functional bias-correction mechanism, leveraging its doubly-robust property to ``debias'' the penalized outcome predictions and ensure that the resulting marginal estimate remains consistent even when the underlying nuisance model is intentionally regularized for stability \citep{chernozhukov2018double, zhang2025bias}.

Even if point estimates are bias-corrected, standard variance estimators for G-computation are often overly optimistic in small samples as a result of overfitting \citep{tackney2023comparison}. A common practical remedy is the ad-hoc small-sample correction proposed by \cite{tsiatis2008covariate}, which scales the variance to account for the degrees of freedom lost to covariates. However, such corrections may be insufficient in high-dimensional regimes where $p$ is large relative to $n$ and/or variable selection is used. 

More rigorous solutions involve sample-splitting and cross-fitting to decouple the nuisance model estimation from the treatment effect calculation \citep{chernozhukov2018double, van2024automated, lee2026improving}. While cross-fitting protects against over-optimistic standard errors, it may suffer from ``data inefficiency'' in small trials where every observation is critical for power. 
To address this, recent work has turned to Higher-Order Influence Functions (HOIF) \citep{robins2016technical, zhao2024covariate}. These estimators can be viewed as versions of standard G-computation that explicitly correct for ``own-observation bias''--the phenomenon where a model overfits to a specific data point and then produces a biased prediction for that same point \citep{robins2016technical, ahrens2025introduction}. By subtracting first and second-order bias terms from G-computation estimators to account for this overfitting, HOIF-based methods provide valid inference even when $p$ is large relative to $n$, effectively correcting the ``excess variance'' that first-order methods ignore. 

The debiased estimators proposed by both \cite{lei2021regression} and \cite{lu2023debiased} align with the principles of the HOIF framework by subtracting first and second-order bias from G-computation estimators, providing a unified theoretical basis for robust covariate adjustment in the absence of cross-fitting. Adopting a finite-population perspective, \cite{lei2021regression} showed that G-computation based on a separate linear outcome model for each treatment arm fitted via OLS incurs a finite-sample bias that can be corrected through higher-order expansion of the bias of the estimator, provided that $\log p$ times the maximum diagonal element of the hat matrix vanishes. Using the same sparsity assumption in \cite{lei2021regression}, \cite{chang2024exact} constructed a G-computation estimator that is exactly unbiased. \cite{lu2023debiased} further improved this approach for linear models to be guaranteed to be at least as efficient as the unadjusted estimator, under the $p=o(n)$ condition. 

Most recently, \cite{zhao2024covariate} demonstrated that the approaches of \cite{lei2021regression}, \cite{lu2023debiased}, and \cite{chang2024exact}, among others \citep{jiang2025adjustments}, can all be mathematically unified under the umbrella of HOIF. Building on this, \cite{abadie2025unbiased} extended the leave-one-out adjustment framework to include ridge regression. By replacing the standard ordinary least squares projection with a ridge-regularized counterpart, they leverage the numerical stability of the penalty to ensure robust performance even as $p$ approaches $n$. While their primary focus remains on the fixed-$p$ case, they show that under the $p=o(n)$ regime, an appropriately tuned ridge penalty achieves efficiency guarantees similar to leave-one-out ordinary least squares without requiring structural sparsity. 

A critical limitation of these corrections is their reliance on the properties of linear projection; as noted by \cite{zhao2026adjusting}, such OLS-based debiasing does not naturally extend to the non-linear link functions used in GLMs. Until recently, this left a significant gap in the theory for binary or count outcomes in high-dimensional trials. \cite{zhao2026adjusting} bridged this gap by utilizing a leave-one-out (LOO) procedure. This Jackknife-based procedure aims to numerically mitigate the same finite-sample biases that \cite{lei2021regression} address analytically, offering a potentially more flexible--albeit computationally intensive-- alternative for non-linear adjustment models. They also introduced a ``calibrated'' extension of this estimator rooted in the ``no-harm'' framework of \cite{cohen2024no}. This approach ensures that the cross-validated predictions are unbiased on average--a necessary condition to guarantee that covariate adjustment does not decrease efficiency relative to the unadjusted estimator. By integrating this calibration step into the Jackknife procedure, \cite{zhao2026adjusting} provides a robust mechanism for non-linear adjustment that maintains valid inference and efficiency parity even as $p$ approaches $n$, though the computational cost of the LOO iterations remains a practical consideration.

\subsection{Main contributions}\label{sec:main_contributions_ch3}
The main contributions of this paper are outlined as follows:
\begin{enumerate}
\item We study if simple approaches can solve the challenges of covariate adjustment in small trials. In particular, we evaluate the performance of variable selection and different variants of cross-fitting in such small sample settings.

\item We explore the use of Bayesian methods with G-computation. Specifically, we apply Bayesian approaches to non-linear models to mitigate problems such as complete separation that commonly occur in small samples. This approach is consistent with the growing regulatory openness to Bayesian methods \citep{us2026use}.

\item We provide accessible insight into more advanced methods, highlighting their connections with simple approaches. We build intuition for why and when they perform well or fail, together with a review of related approaches in the literature.

\item We conduct independent simulation studies to evaluate the performance of methods across a range of settings that vary the ratio of the number of covariates to the sample size. For example, our simulation scenarios cover covariate-to-sample size ratios observed in the 50 RCTs analyzed by \cite{shao2026benchmarking}. Furthermore, we consider both parametric simulations, which allow us to explore a wide variety of data-generating mechanisms, and resampling from real datasets, which reflects practically relevant mechanism for the study at hand \citep{morris2019using}.
\end{enumerate}

\subsection{Notation and estimand}\label{sec:notation_estimand_ch3}
Consider an RCT that yields independent and identically distributed (i.i.d.) observations from a sample of $n$ individuals. For each participant $i=1,\ldots,n$, the observed data consist of a $p$-dimensional vector of baseline covariates $X_i \in \mathbb{R}^p$, measured prior to treatment assignment, a binary treatment assignment indicator $A_i \in \{0,1\}$, and the primary outcome of interest $Y_i$. The indicator $A_i=1$ indicates assignment to the active treatment arm, whereas $A_i=0$ indicates assignment to the control arm. The observed dataset can therefore be written as $\{(X_i, A_i, Y_i)\}_{i=1}^n$, which we regard as i.i.d.\ realizations from the joint distribution of $(X, A, Y)$.

We assume that treatment is allocated by simple randomization, independently of baseline covariates. Let $\pi = \Pr(A_i = 1)$ denote the fixed randomization probability, that is, the probability that an individual is assigned to the active treatment arm. Let $n_1 = \sum_{i=1}^n A_i$ denote the number of individuals assigned to the active treatment arm and $n_0 = n-n_1$ the number assigned to the control arm. 

We focus on the average treatment effect (ATE), which is often the primary target of interest in clinical trials \citep{ICH2020}. The ATE is defined as: 
\begin{align}\label{eq:ATE}
   \tau=\E\left(Y^1-Y^0\right),
\end{align}
where $Y^1$ and $Y^0$ denote the potential (or counterfactual) outcomes that would have been observed under assignment to treatment and control, respectively. Alternative causal estimands, including the risk ratio and odds ratio, may also be targeted depending on the study objectives.

\subsection{G-computation estimators}\label{sec:stand_G_comp_ch3}
To estimate the ATE $\tau$, we employ G-computation. This approach begins by specifying a working model for the conditional mean of the outcome given treatment and baseline covariates, $\E(Y|A, X) = m(A, X; \theta)$, where $m(\cdot)$ is a known functional form indexed by unknown parameters $\theta$. Common specifications include the linear model $m(A, X; \alpha, \beta, \gamma) = \alpha + \beta A + \gamma'X$ for continuous outcomes and the logistic model $m(A, X; \alpha, \beta, \gamma) = \text{expit}(\alpha + \beta A + \gamma'X)$ for binary outcomes. Here, $X$ may include any transformations of the covariates.

After fitting the model (e.g., via maximum likelihood), we obtain predicted potential outcomes for each participant as $\hat{m}^{a}(X_i) = m(A=a, X_i; \hat{\theta})$ for $a \in \{0, 1\}$. For instance, in the linear case, the predictions are $\hat{m}^{1}(X_i) = \hat{\alpha} + \hat{\beta} + \hat{\gamma}'X_i$ and $\hat{m}^{0}(X_i) = \hat{\alpha} + \hat{\gamma}'X_i$. The G-computation estimator of $\tau$ is then obtained by averaging the difference between these predictions across the sample:
\begin{align}\label{eq:hat_tau_gcomp}
    \hat{\tau}_{\text{GComp}} = \frac{1}{n} \sum_{i=1}^{n} \left\{ \hat{m}^{1}(X_{i}) - \hat{m}^{0}(X_{i}) \right\}.
\end{align}

When based on canonical GLMs fitted via MLE, it is numerically equivalent to the augmented inverse probability weighting (AIPW) estimator based on estimated marginal randomization probabilities $\hat{\pi} = n_1/n$:
\begin{align*} 
\hat{\tau}_{\text{AIPW}} = \frac{1}{n}\sum_{i=1}^{n} \left[ \left( \frac{A_i}{\hat{\pi}} \{Y_i-\hat{m}^{1}(X_{i})\} + \hat{m}^{1}(X_{i}) \right) - \left( \frac{1 - A_i}{1 - \hat{\pi}} \{Y_i-\hat{m}^{0}(X_{i})\} + \hat{m}^{0}(X_{i}) \right) \right]. 
\end{align*}
This equivalence stems from the score equations of canonical GLMs, which impose ``prediction unbiasedness'' constraints $\sum_{i=1}^{n} A_i \{Y_i - \hat{m}^1(X_i)\} = 0$ and $\sum_{i=1}^{n} (1 - A_i) \{Y_i - \hat{m}^0(X_i)\} = 0$ \citep{bannick2025general}.
A crucial consequence of this AIPW representation is that the G-computation estimator remains consistent for the ATE even under misspecification of the outcome model \citep{tsiatis2008covariate, diaz2016enhanced}. This inherent robustness, which is generally not shared by the conditional MLE of the treatment effect coefficient itself (with the exception of linear models), makes G-computation highly appealing for regulatory and clinical trial applications \citep{FDA2023}. While our primary focus is on the G-computation estimator, we utilize this AIPW representation for notational and analytical convenience in our subsequent theoretical derivations.

\section{Problems of overfitting}\label{sec:problems_of_over_fitting_ch3}
GLMs are prone to overfitting when the number of covariates is large relative to the sample size, in which case they may capture noise rather than the true signal \citep{montesinos2022multivariate}. Logistic regression is especially vulnerable to this, as we illustrated in \autoref{figure:dist_MLE_bin}. In such settings, the MLEs may even fail to exist due to complete separation \citep{stolte2024comprehensive}. 

While these problems have been well studied for the MLE \citep{portnoy1988asymptotic,cordeiro1991bias,sur2019modern}, this is much less the case for G-computation estimators of the treatment-specific outcome mean in randomized experiments (see \cite{yadlowsky2021sloe} for an exception). These have much less stringent requirements as they are root-$n$ consistent so long as the outcome predictions $\hat{m}^a(X)$ converge to some limit $m^{a*}(X)$ for $a=0,1$, no matter at what rate \citep{van2024automated}, in the sense that 
\begin{equation}\label{2nd}
   \left(\frac{\pi}{\hat{\pi}}-1\right)\E\left\{\hat{m}^a(X)-m^{a*}(X)\right\}=o_{p}({n^{-1/2}}).
\end{equation}
While this requirement is very weak, it may fail when the number of covariates is large relative to the sample size, as MLEs may then fail to converge \citep{portnoy1988asymptotic} and, consequently, the mean squared error $\E\left[\left\{\hat{m}^a(X)-m^{a*}(X)\right\}^2\right]$ (which upper bounds $|\E\left\{\hat{m}^a(X)-m^{a*}(X)\right\}|$) may not converge to zero (see Proposition 1 in \cite{yadlowsky2021sloe} for a formal proof in the study of proportional asymptotic regimes, where the number of covariates grows proportionally with the sample size, and Section \ref{appe_G_computation_with_many_covariates_ch3} for additional detail). In view of this, we will study the impact of a number of simple, off-the-shelf solutions to accommodate this in small sample settings. We will contrast these with more advanced techniques later in the paper.

\subsection{Variable selection}\label{sec:variable_selection_ch3}
Data-adaptive variable selection is widely used to mitigate overfitting when the number of covariates is large relative to the sample size (e.g., \cite{tibshirani1996regression,van2008high, belloni2009least,wager2016high,xia2023debiased}). By eliminating noise variables, these methods reduce model complexity and improve estimation stability. While this in turn introduces regularization bias, condition (\ref{2nd}) suggests that the impact of this may be minor. In this study, we focus on LASSO-based variable selection, as well as post-LASSO, which is even less affected by regularization bias. For simplicity, we describe the procedure for a linear outcome model, although an analogous procedure can be applied for logistic outcome models:
\begin{enumerate}
\item Estimate $\Gamma=\{\alpha,\beta,\gamma\}$ (for the linear model described in Section \ref{sec:stand_G_comp_ch3}) using the Lasso estimator:
\begin{align*}
    \widehat{\Gamma}^{\text{Lasso}} = \arg\min_{\Gamma \in \mathbb{R}^{p+2}} \left\{ \frac{1}{n} \sum_{i=1}^n \left(Y_i-\Gamma'Z_{i}\right)^2 + \lambda \sum_{j=1}^{p+2} |\Gamma_j| \right\},
\end{align*}
where $Z_i=\{1,A_i,X_i\}$ and $\lambda>0$ is a regularization parameter.
\item Post-Lasso estimator: let $S = \left\{ j \in \{1, \dots, p+2\} : \widehat{\Gamma}_j^{\text{Lasso}} \neq 0 \right\}$ denote the set of selected variables. To reduce shrinkage bias,  we refit the model using MLE only on the variables selected by Lasso:
\begin{align*}
    \widehat{\Gamma}^{\text{post-Lasso}} = \arg\min_{\Gamma \in \mathbb{R}^{|S|}} \left\{ \frac{1}{n} \sum_{i=1}^n \left(Y_i-\Gamma'Z_{i,S}\right)^2 \right\},
\end{align*}
where $Z_{i,S}$ denotes the sub-vector of $Z_i$ indexed by $S$. We ensure that the treatment variable is always included in the model by assigning it a penalization factor of zero in the Lasso step.
\item Compute $\hat{m}_{\text{Lasso}}^{a}(X)$ for $a \in \{0,1\}$, based on $\widehat{\Gamma}^{\text{post-Lasso}}$.
\item The post-Lasso G-computation estimator is then given by 
\begin{align*}
    \hat{\tau}_{\text{Lasso}}=\frac{1}{n}\sum_{i=1}^{n}\left\{\hat{m}_{\text{Lasso}}^{1}(X_i)-\hat{m}_{\text{Lasso}}^{0}(X_i)\right\}.
\end{align*}
\end{enumerate}
To the best of our knowledge, there is no formal study of G-computation after use of variable selection procedures in settings where the usual sparsity assumptions fail.

\subsection{Firth correction and Bayesian methods}\label{sec:firth_correction_ch3}
As discussed in Section \ref{sec:problems_of_over_fitting_ch3}, overfitting can lead to situations in which the MLE does not exist. This may arise in logistic regression, for example, when the covariates perfectly distinguish the two outcome classes so that there is complete separation. In such cases, the MLEs of the regression coefficients diverge \citep{albert1984existence}. Firth correction mitigates the small-sample bias in MLEs by removing the first-order bias from the score equation, thereby addressing the problem of complete separation \citep{firth1993bias,varewyck2014shrinkage, puhr2017firth, xu2026using}. For logistic regression, this is available in \texttt{R} via the \texttt{logistf} function from the \texttt{logistf} package. We will study if standard  Firth correction mitigates the bias affecting G-computation estimators in small RCTs. We will additionally consider Bayesian logistic regression with weakly informative priors, such as the Cauchy prior distribution \citep{gelman2008weakly}. This also addresses complete separation and is available in \texttt{R} using the \texttt{bayesglm} function from the \textit{arm} package.

Although these methods may mitigate the issue of overfitting and complete separation, they may also compromise the robustness of the G-computation estimator by violating the prediction unbiasedness condition. To restore this condition, we employ targeted maximum likelihood estimation (TMLE) procedures \citep{gruber2010targeted}. Specifically, we fit a standard logistic regression model that includes an intercept, treatment indicator, and an offset given by the logit of the initial outcome model predictions obtained from either Firth or Bayesian GLM. The model can be written as
\begin{align*}
    \E(Y=1 \mid A, X) = \logit^{-1}\left[\varepsilon_0 + \varepsilon_1 A + \logit\left\{\hat{m}(A,X)\right\}\right],
\end{align*}
and is fitted using maximum likelihood. Here, $\hat{m}(A,X)$ denotes the predicted outcome (based on both treatment and covariates) from either the Firth or the Bayesian logistic model. Using the estimated fluctuation parameters, $\hat{\varepsilon}_0$ and $\hat{\varepsilon}_1$, we obtain updated predictions of the outcome under treatment and control. For individual $i$, these are given by $m^{\ast1}(X)=\logit^{-1}\left[\logit\left\{\hat{m}^1(X)\right\}+\hat{\varepsilon}_0+\hat{\varepsilon}_1\right]$ and $m^{\ast0}(X)=\logit^{-1}\left[\logit\left\{\hat{m}^0(X)\right\}+\hat{\varepsilon}_0\right]$. The updated predictions satisfy the prediction unbiasedness condition defined in Section \ref{sec:stand_G_comp_ch3}, as this is guaranteed by the score equations for $\hat{\varepsilon}_0$ and $\hat{\varepsilon}_1$ under canonical GLMs.

\section{Underestimation of standard errors}\label{sec:underestimation_of_SE_ch3}
The variance of the G-computation estimator \eqref{eq:hat_tau_gcomp} using canonical GLMs is commonly estimated as $1/n$ times the sample variance of the estimated influence functions:
\begin{align}\label{eq:var_inf}
	\frac{A_{i}}{\hat{\pi}}\left\{Y_{i}-\hat{m}^{1}(X_{i})\right\}+\hat{m}^{1}(X_{i})
	-\frac{1-A_{i}}{1-\hat{\pi}}\left\{Y_{i}-\hat{m}^{0}(X_{i})\right\}-\hat{m}^{0}(X_{i}),
\end{align}
where $\hat{m}^{1}(X_{i})$ and $\hat{m}^{0}(X_{i})$ denote the predicted outcomes under treatment and control obtained using the standard procedures described in Section \ref{sec:stand_G_comp_ch3}. When the outcome model is overfitted, the residual terms $Y_i-\hat{m}^{1}(X_i)$ and $Y_i-\hat{m}^{0}(X_i)$ may become artificially small. Consequently, the first and third terms in Equation \eqref{eq:var_inf} may be systematically too close to zero, leading to an underestimation of the variance and, therefore, the standard errors. This phenomenon is illustrated in \autoref{figure:dist_G_Comp_bin}, which shows the distribution of the estimated standard errors together with the Monte Carlo standard deviation of the estimated ATE. The problem observed in this figure is the result of so-called own-observation bias, which occurs when the same observations are used both to fit the outcome model and to evaluate the treatment effect. 
Although variable selection procedures may partially address this issue, small sample corrections or sample splitting are specifically designed to mitigate own-observation bias.

\subsection{Small-sample correction}\label{sec:small_sample_correction_ch3}
Small-sample corrections to the variance estimator \eqref{eq:var_inf}, for instance by multiplying it by a factor $\frac{n-1}{n-p-1}$  \citep{tsiatis2008covariate} can partially address the own-observation bias affecting the variance estimator. This factor replaces the default $n-1$ effective degrees of freedom correction for the empirical variance estimator by $n-p-1$ to account for the estimation of $p$ parameters in the outcome model. However, this correction has no formal justification outside of ordinary least squares regression for the outcome, and has no immediate extension to other estimation strategies (e.g., LASSO or post-LASSO).

\subsection{Cross-fitting}\label{sec:cross_fitting_ch3}
Cross-fitting is widely used to avoid using the same observation both to estimate nuisance functions and to evaluate the target estimator, an issue that is particularly important for variance estimation \citep{chernozhukov2018double,van2024automated,smith2025performance}. To illustrate, consider the simple problem of estimating a population variance. The empirical variance computed around the sample mean is biased downward because the same observations are used both to estimate the mean and to compute the squared deviations, requiring a degrees-of-freedom correction.  Cross-fitting avoids this reuse of data, often referred to as own-observation bias, by evaluating each observation using a mean estimated from other observations.

By the same logic, cross-fitting may also benefit variance estimation of the G-computation estimator. It prevents the variance of the residual terms, $Y_i-\hat{m}^{1}(X_i)$ and $Y_i-\hat{m}^{0}(X_i)$ in Equation \eqref{eq:var_inf}, from being systematically underestimated, thereby reducing the need for small-sample corrections to the standard errors. It is therefore of interest to study whether cross-fitting improves G-computation estimators in settings with more complex nuisance structures. In particular, we examine whether cross-fitting remains effective when the number of covariates is large relative to the sample size, a regime for which formal theory is currently limited. We study this more formally in Section \ref{sec:leave_one_out_ch3}.

Cross-fitting based G-computation estimators can be implemented as follows:
\begin{enumerate}
\item Split the data into $K$ mutually exclusive folds ($\mathcal{I}_{k}, k=1,\dots,K$).
\item For each fold $k$:
\begin{enumerate}
    \item Estimate the parameter vector \(\Gamma=\{\alpha,\beta,\gamma\}\) (for example, for the linear model described in Section \ref{sec:stand_G_comp_ch3}) by maximum likelihood using the observations in the complement set \(\mathcal{I}_k^c=\{1,\dots,n\}\setminus \mathcal{I}_k\). 
\item Using the fitted model, compute the predicted outcomes \(\hat{m}_k^a(X_i)\) for \(a\in\{0,1\}\) for each individual \(i\in \mathcal{I}_k\).
\end{enumerate}
\item The cross-fitting G-computation estimator is then
\begin{align*}
    \hat{\tau}_{\text{CF}}=\frac{1}{n}\sum_{k=1}^{K}\sum_{i\in\mathcal{I}_k}\left\{\hat{m}^{1}_{k}(X_i)-\hat{m}^{0}_{k}(X_i)\right\}.
\end{align*}
\end{enumerate}

One can also consider the cross-fitted AIPW estimator,
\begin{align*}
 \frac{1}{n}\sum_{k=1}^{K}\sum_{i\in\mathcal{I}_k}\left[\frac{A_{ik}}{\hat{\pi}_{k}}\left\{Y_{ik}-\hat{m}^{1}_{k}(X_i)\right\}+
    \hat{m}^{1}_{k}(X_i)-\frac{1-A_{ik}}{1-\hat{\pi}_{k}}\left\{Y_{ik}-\hat{m}^{0}_{k}(X_i)\right\}-\hat{m}^{0}_{k}(X_i)\right],
\end{align*}
which is better suited to cross-fitting than the G-computation estimator as the residual terms directly correct for the finite-sample error introduced when nuisance functions are estimated on held-out folds. Here, the randomization probability can also be estimated over all folds. 

\subsection{Leave-one-out cross-fitting based estimators}\label{sec:leave_one_out_ch3}
Recently, leave-one-out cross-fitting has been proposed as a variant of standard cross-fitting to mitigate own-observation bias \citep{zhao2024covariate,zhao2026adjusting,lee2026improving}. This is of interest to make maximum use of the available information in small trials and has the added advantage of being deterministic (i.e., not dependent on a random choice of folds), thus not being susceptible to simulation error. For instance, for the mean outcome under treatment, the leave-one-out cross-fitted AIPW estimator can be defined as
\begin{align*}
  \frac{1}{n}\sum_{i=1}^{n} \left\{\frac{A_{i}}{\hat{\pi}}Y_{i}-\left(\frac{A_{i}}{\hat{\pi}}-1\right)\hat{m}^{1,-i}(\tilde{X}_i)\right\},
\end{align*}
where $\hat{m}^{1,-i}(\tilde{X}_i) = m(\tilde{X}_i;\hat{\delta}_1^{-i})$ is the leave-one-out predicted outcome from a model of $Y$ on $\tilde{X}$ (a covariate vector with an intercept) fitted in the treated group, with the leave-one-out estimates of regression coefficient $\hat{\delta}_1^{-i}$ defined as a solution to
\begin{align}\label{eq:score}
  \frac{1}{n-1} \sum_{j \neq i} \tilde{X}_j \frac{A_j}{\hat{\pi}}\left\{Y_j - m(\tilde{X}_j;\hat{\delta}_1^{-i})\right\} = 0.
\end{align}
The form of $m(\tilde{X}_i; \hat{\delta}_1^{-i})$ depends on the model. For example, it equals $\logit^{-1}(\tilde{X}_i^\top \hat{\delta}_1^{-i})$ for the logistic model, and $\tilde{X}_i^\top \hat{\delta}_1^{-i}$ for the linear model.

Since we are considering a setting where treatment and covariates are independent, the second term in Equation (\ref{eq:score}) has the same mean as $\frac{1}{n} \sum_{i=1}^n \Tilde{X}_im(\Tilde{X}_j;\hat{\delta}_1^{-i})$, which makes more efficient use of the data \citep{zhao2026adjusting}. As a result, $\hat{\delta}_1^{-i}$ can alternatively be solved from
\begin{align}\label{eq:score_equation_JASA}
  \frac{1}{n - 1} \sum_{j \ne i} \Tilde{X}_j\frac{A_j}{\hat{\pi}} Y_j-\frac{1}{n} \sum_{l=1}^{n} \Tilde{X}_l\, m\left(\Tilde{X}_l;\hat{\delta}_1^{-i}\right)= 0.
\end{align}
The AIPW estimator based on it, is called a jackknife score-based adjusted (JASA) estimator \citep{zhao2026adjusting}, denoted by $\hat{\tau}_{\text{JASA}}$. It accommodates canonical GLMs and is $\sqrt{n}$-consistent and asymptotically normal when $p = o(n^{2/3})$. 

For a linear model $m(\Tilde{X}_i;\hat{\delta}_1^{-i})=\Tilde{X}_i^\top \hat{\delta}_1^{-i}$ fitted in the treatment group, the JASA estimator uses
\begin{align*}
\hat{\delta}_1^{-i}=\frac{n}{n-1}\left(\Tilde{X}^\top \Tilde{X}\right)^{-1}\sum_{j\neq i} \Tilde{X}_j \frac{A_j}{\hat{\pi}} Y_j.
\end{align*}
The corresponding leave-one-out prediction $\hat{m}^{1,-i}(\Tilde{X}_i)$ can thus be written as
\begin{align*}
&\frac{n}{n-1}\sum_{j\neq i}H_{ij}\frac{A_j}{\hat{\pi}}Y_j=\frac{n}{n-1}\left(\sum_{j=1}^{n} H_{ij} \frac{A_j}{\hat{\pi}} Y_j-H_{ii}\frac{A_i}{\hat{\pi}}Y_i\right),
\end{align*}
where $H_{ij}=\Tilde{X}_i^\top (\Tilde{X}^\top \Tilde{X})^{-1}\Tilde{X}_j$ denotes the $(i,j)$th element of the hat matrix. Upon ignoring the fraction $n/(n-1)$, the AIPW estimator based on these predictions can be written as
\begin{equation}\label{eq:tau_HOIF_1}
\begin{aligned}
\hat{\tau}^\dagger_{\text{HOIF}}
& = \frac{1}{n}\sum_{i=1}^{n}\left\{\frac{A_{i}}{\hat{\pi}}Y_{i}
-\left(\frac{A_{i}}{\hat{\pi}}-1\right)
\left(\sum_{j=1}^{n} H_{ij} \frac{A_j}{\hat{\pi}}Y_j-H_{ii}\frac{A_{i}}{\hat{\pi}}Y_{i}\right)\right\},
\end{aligned}
\end{equation}
which equals a higher-order influence function (HOIF)-based estimator \citep{zhao2024covariate}. This shows that JASA estimators are closely related to HOIF-based estimators, extending them beyond the linear model. This approximate equivalence makes them particularly appealing, as HOIF-based estimators correct for higher-order bias terms that may be non-negligible when the number of covariates is large relative to the sample size (see the next paragraph for more detail).

In Section \ref{appe_leave_one_out_ch3}, we give more insight into the role of leave-one-out cross-fitting in removing own-observation bias. Here, we summarize the key results. In particular, the conditional expectation of the second term in Equation \eqref{eq:tau_HOIF_1}, conditional on $\sum_{k=1}^nA_{k}$ to guarantee that $\hat{\pi}$ can be treated as fixed, equals
 \begin{align}\label{eq:bias_HOIF}
    \frac{1}{n\hat{\pi}^2} \sum_{i,j:i\ne j}\E\left( H_{ij}Y_j^1\right)\E\left(A_iA_j\,\middle|\, \sum_{k=1}^nA_{k}=n_1\right)-
    \frac{1}{n\hat{\pi}} \sum_{i,j:i\ne j}\E\left( H_{ij}Y_j^1\right)\E\left(A_j\,\middle|\, \sum_{k=1}^nA_{k}=n_1\right),
 \end{align}
where we use the i.i.d. observations together with the randomization condition that treatment assignment is independent of baseline covariates.
As shown in Section \ref{appe_leave_one_out_ch3}, Equation \eqref{eq:bias_HOIF} simplifies to
\begin{align*}
    \frac{1-\hat{\pi}}{\hat{\pi}(n-1)}\E\left\{H_{ii}\E\left(Y_i^1|\Tilde{X}\right)\right\}.
\end{align*}
Under standard regularity conditions on the observed data, in particular that the conditional mean of the potential outcome is bounded, it follows from this that the bias of $\hat{\tau}^\dagger_{\text{HOIF}}$ is of order $O\left(\frac{n_0}{n_1}\frac{p}{n^2}\right)$. This bias is smaller than that of the standard AIPW estimator $\hat{\tau}_{\text{AIPW}}$, which can be shown (by a similar argument) to be of order $O\left(\frac{n_0}{n_1}\frac{p}{n}\right)$; see Section \ref{appe_leave_one_out_ch3}. It can be attenuated further by centering the outcome in the second term in Equation \eqref{eq:tau_HOIF_1} as follows:
\begin{equation}\label{eq:tau_HOIF_c}
\begin{aligned}
\hat{\tau}^\ddagger_{\text{HOIF}}& = \frac{1}{n}\sum_{i=1}^{n} \left\{\frac{A_{i}}{\hat{\pi}}Y_{i}-\left(\frac{A_{i}}{\hat{\pi}}-1\right)\left(\sum_{j=1}^{n} H_{ij} \frac{A_j}{\hat{\pi}}(Y_j-C_{1})-H_{ii}\frac{A_{i}}{\hat{\pi}}(Y_i-C_{1})\right)\right\},
\end{aligned}
\end{equation}
where $C_1= \frac{\sum_{i=1}^{n}A_iH_{ii}Y_i}{\sum_{i=1}^{n}A_iH_{ii}}$. These centering terms deliver an estimator with similar performance, as demonstrated in simulations, as that proposed by \cite{zhao2024covariate}, who use $C_1 = \frac{1}{n} \sum_{i=1}^{n} \frac{A_i}{\hat{\pi}} Y_i$. The proposed centering terms make the second term in Equation \eqref{eq:tau_HOIF_c} exactly mean zero, meaning they will not introduce bias into the estimator (see Section \ref{appe_leave_one_out_ch3}).

To reduce the variance of $\hat{\tau}_{\text{JASA}}$ in settings with a large number of covariates, \cite{zhao2026adjusting} propose a post-hoc  calibration step based on the initial leave-one-out predictions. For each treatment group $a \in \{0,1\}$, the calibration coefficients are defined by
\begin{align*}
(\hat{\eta}_a, \hat{\eta}_{a,0}, \hat{\eta}_{a,1}):= \arg\min_{\eta_a, \eta_{a,0}, \eta_{a,1}} 
\sum_{i:A_i = a} \left\{ Y_i-\eta_a - \eta_{a,0}\,\hat{m}^{0,-i}(\Tilde{X}_i) - \eta_{a,1}\,\hat{m}^{1,-i}(\Tilde{X}_i) \right\}^2.
\end{align*}
As before, $\hat{m}^{1,-i}(\tilde{X}_i)$ denotes the leave-one-out predicted outcome under treatment, computed based on the leave-one-out regression estimate $\hat{\delta}_1^{-i}$ that solves \eqref{eq:score_equation_JASA}. Similarly, $\hat{m}^{0,-i}(\tilde{X}_i)$ denotes the corresponding leave-one-out predicted outcome under control, obtained upon replacing $A$ by $1-A$. The calibrated prediction for unit $i$ under treatment $a$ is then
\begin{align*}
\Tilde{m}^a(\Tilde{X}_i) = \hat{\eta}_a + \hat{\eta}_{a,0}\,\hat{m}^{0,-i}(\Tilde{X}_i) + \hat{\eta}_{a,1}\,\hat{m}^{1,-i}(\Tilde{X}_i),
\end{align*}
and the calibrated JASA estimator for the mean outcome under treatment is
\begin{align*}
   \hat{\tau}_{\text{JASA\_cal}}& = \frac{1}{n}\sum_{i=1}^{n} \left\{\frac{A_{i}}{\hat{\pi}}Y_{i}-\left(\frac{A_{i}}{\hat{\pi}}-1\right)\Tilde{m}^1(\Tilde{X}_i)\right\}.
\end{align*}
We will evaluate whether these estimators outperform standard cross-fitting based estimators, particularly in small sample settings with a relatively large number of covariates.

 \section{Simulation studies}\label{sec:simulations_ch3}
\subsection{Data generating mechanisms}
\subsubsection{Continuous outcomes}
We consider three settings for continuous outcomes. Setting 1 follows the data-generating mechanism of \cite{zhao2024covariate}, but with infinite (instead of finite) population. A vector of baseline covariates $X$ is generated from a multivariate normal distribution, $\text{Normal}(0, \Sigma)$, where the covariance matrix satisfies $\Sigma_{ij}=0.1^{|i-j|}, i,j=1\dots p$. The treatment indicator $A$ is generated from a Bernoulli distribution with probability $0.5$, i.e., $A \sim \text{Bernoulli}(0.5)$. The outcome $Y$ is generated as \[Y =A+\gamma'X+\epsilon \cdot\sqrt{\text{Var}(\gamma'X)/\text{Var}(\epsilon)}/\sqrt{\delta},\] where the covariate coefficients are defined as $\gamma_j = \frac{(-1)^j}{\log(j+1)}, j = 1, \dots, p$ and $\epsilon \sim \text{Normal}(0,1)$. The parameter $\delta$ controls the signal-to-noise ratio and is fixed at $\delta=2$.

In the second setting (Setting 2), we consider the same data-generating mechanism but impose a sparsity assumption on the coefficient vector $\gamma$, with $\gamma_j \neq 0$ for $j=1,\dots,5$ and $\gamma_j = 0$ for $j=6,\dots,p$.

The third setting (Setting 3) is designed to mimic the BestAIR trial \citep{zhao2017effect}. Baseline covariates are generated by resampling patients with replacement, and the treatment indicator is generated from a Bernoulli distribution with probability 0.5. 
The outcome $Y$ (change in mean 24-hour systolic blood pressure at six months) is generated via the linear main effects model:
\[Y =126.2-2.5A+\gamma'X+\epsilon \cdot\sqrt{\text{Var}(\gamma'X)/\text{Var}(\epsilon)}/\sqrt{\delta},\]
where $\epsilon\sim \text{Normal}(0,6.4)$. Here, all coefficients (including $\gamma$ whose specific values are provided in \autoref{table:appe_gamma_setting3_continuous_ch3} in Section \ref{appe_sim_results_ch3}) and the variances $\text{Var}(\gamma^\top X)$ and $\text{Var}(\epsilon)$ are estimated on the BestAIR trial data, with $\delta = \text{Var}(\gamma^\top X)/\text{Var}(\epsilon)$ estimated as 2.86.

We consider several aspect ratios $k$, defined as the ratio of the number of covariates to the sample size. Specifically, $k$ = 0.05, 0.1, 0.2, 0.3, 0.4 and 0.7 for Settings 1 and 2. For Setting 3, we consider $k$ = 0.05, 0.1, 0.2, 0.3, 0.4 and 0.54, where the largest value reflects the maximum covariate-to-sample size ratio after excluding covariates with missing values in the BestAIR trial. Sample sizes of 50, 100 and 200 are considered for Setting 1 and Setting 2, while a sample size of 124 (corresponding to the number of patients in the BestAIR trial after excluding those with missing outcomes) is used for Setting 3. For each value of $k$ in Setting 3, we include the first $k \times n$ covariates from the BestAIR data set, whose names are listed in \autoref{figure:appe_covariates_names_ch3} in Section \ref{appe_sim_results_ch3}.

\subsubsection{Binary outcomes}\label{subsec:DGM_binary_outcomes}
We likewise consider three settings for binary outcomes. In Setting 1, baseline covariates and treatment are generated as in Setting 1 of the previous section, and the outcome as $Y \sim \text{Bernoulli}\left\{\mu(A, X)\right\}$, where $$\mu(A, X)= \text{logit}^{-1}(0.25 + 0.75A+\gamma'X)$$ with covariate coefficients defined as $\gamma_j = \frac{(-1)^j}{\sqrt{j}}, j = 1, \dots, p.$ 
Setting 2 imposes the same sparsity assumption as in the case of continuous outcomes.
Finally, in Setting 3, baseline covariates and treatment are generated as in Setting 3 of the previous section and the outcome (a binary indicator of high systolic blood pressure $(\geq130 mmHg)$ at six months) is modeled as $Y \sim \text{Bernoulli}\left\{\mu(A, X)\right\}$ with
\begin{align*}
    \mu(A, X)= \text{logit}^{-1}(-1.9-1.1A+\gamma'X).
\end{align*}
with all coefficients (including $\gamma$ whose specific values are provided in \autoref{table:appe_gamma_setting3_binary_ch3} in Section \ref{appe_sim_results_ch3}) estimated on the BestAIR trial data. We consider the aspect ratios and sample sizes as in the continuous outcome setting.

\subsection{Estimands, estimators and performance measures}\label{subsec:estimators_and_performance_measures_ch3}
For continuous outcomes, the ATE is 1 in Settings 1 and 2, and $-2.5$ in Setting 3, as determined by the data generating mechanisms. For the binary outcomes, the ATE is computed as the difference between the average inverse-logit values under treatment and under control across $10^6$ observations. Because the logistic link is non-collapsible, the ATE varies across settings and covariate-to-sample size ratios (see \autoref{table:bin_s1_bias_SD_SE_COV} and \autoref{table:appe_bin_s2_bias_SD_SE_COV_ch3}).

For continuous outcomes, we fit a linear regression with intercept, treatment indicator, and main effects of covariates. We evaluate standard G-computation (Section \ref{sec:stand_G_comp_ch3}), post-LASSO G-computation (Section \ref{sec:variable_selection_ch3}), cross-fitted G-computation (Section \ref{sec:cross_fitting_ch3}), and cross-fitting combined with post-LASSO G-computation (see Section \ref{appe_cross_fitting_variable_selection_ch3}). For cross-fitted estimators, we use 5-fold sample splitting. In addition, we consider the JASA and HOIF-based estimators described in Section \ref{sec:leave_one_out_ch3}. For comparison, we also include unadjusted estimators.

For binary outcomes, we consider the same estimators, but now based on Bayesian logistic regression with a Cauchy prior distribution, in addition to Firth-based G-computation (see Section \ref{sec:firth_correction_ch3}). For the latter, we fit a Firth-corrected logistic model with an intercept, a treatment indicator, and main effects covariates. As discussed in Section \ref{sec:firth_correction_ch3}, a TMLE step is applied to all Bayesian and Firth-based G-computation estimators.

The variance of each estimator is computed using the influence function, specifically as $1/n$ times the sample variance of the estimated influence function, as shown in Equation \eqref{eq:var_inf} for the standard G-computation estimator. The same formula is applied to the other estimators, with the standard outcome predictions in Equation \eqref{eq:var_inf} replaced by the predictions obtained from the corresponding estimator. For HOIF-based estimators, the variance is computed likewise, but treated and control components are each centered by subtracting their respective group-specific means (see Section \ref{appe_variance_estimation_HOIF_ch3}). We also consider small sample corrections for the standard and post-LASSO G-computation estimators for both continuous and binary outcomes (see Section \ref{sec:small_sample_correction_ch3}). For continuous outcomes, we additionally evaluate the HC1 and HC3 sandwich variance estimators (see Section \ref{appe_sandwich_variance_ch3}) in the context of main-effects linear outcome models, where the G-computation estimator coincides with the estimated treatment coefficient, i.e., $\hat{\tau}_{\text{GComp}} = \hat{\beta}$.

For each estimator and simulation setting, we report the estimated bias, Monte Carlo standard deviation (SD), average standard error (SE), and the coverage probability (COV) of the 95\% confidence intervals (obtained as estimate $\pm 1.96$ standard error) for the ATE, based on 1000 simulation runs.
Bias reflects systematic error, the Monte Carlo SD shows empirical variability, the average SE assesses the accuracy of variance estimation, and coverage probability evaluates whether the resulting 95\% confidence intervals attain their nominal coverage level.

\subsection{Results}
\subsubsection{Continuous outcomes}
Results for all settings are reported in \autoref{table:cont_s1_bias_SD_SE_COV} for Setting 1, \autoref{table:appe_cont_s2_bias_SD_SE_COV_ch3} in Section \ref{appe_sim_results_ch3} for Setting 2, and \autoref{table:cont_s3_bias_SD_SE_COV} for Setting 3. Across all settings, the estimators are generally unbiased and perform similarly; the HOIF-based estimators with uncentered outcomes appear biased at higher values of $k$ in Setting 3, but this is often on par with the unadjusted estimator, thus reflecting the increased outcome variability at higher $k$. In particular, the conditional mean of the outcome is unbounded as $k$ grows in this setting, unlike in Settings 1 or 2. 

Standard G-computation estimators, followed by post-LASSO G-computation, yield the least reliable variance estimates, with coverage dropping sharply as $k$ increases. This is the result of own-observation bias, failure of the ultra-sparsity assumptions on which the LASSO relies  \citep{belloni2014inference,bradic2019sparsity}, as well as LASSO's tendency to select more covariates than the number of covariates with non-zero coefficients in the model; see \autoref{figure:appe_cont_s2_p_selected_ch3} in Section \ref{appe_sim_results_ch3}. All other estimators (which are based on cross-fitting) come with valid standard errors, though these tend to be conservative at large $k$ when cross-fitting is used without LASSO, in HOIF-based estimators with uncentered outcomes or JASA estimators without calibration. These estimators also tend to be less precise. The remaining estimators (cross-fitting combined with post-LASSO, calibrated JASA and HOIF-based estimators with centered outcomes) are competitive and deliver good coverage, with the latter generally performing best.

\begin{table}[h!]
    \caption{Estimated bias, Monte Carlo standard deviation, average standard errors, and coverage probabilities for ATE obtained from Setting 1 with continuous outcomes.}
    \begin{center}
 \resizebox{\textwidth}{!}{
\begin{tabular}{ccccccccccccccccccc}
\toprule
$k$ & $\hat{\tau}_{\text{\tiny{Unadj}}}$ & $\hat{\tau}_{\text{\tiny{GComp}}}$ & $\hat{\tau}_{\text{\tiny{Lasso}}}$ & $\hat{\tau}_{\text{\tiny{CF}}}$ & $\hat{\tau}_{\text{\tiny{CF\_Lasso}}}$ & $\hat{\tau}^\dagger_{\text{\tiny{HOIF}}}$ & $\hat{\tau}^\ddagger_{\text{\tiny{HOIF}}}$ & $\hat{\tau}_{\text{\tiny{JASA}}}$ & $\hat{\tau}_{\text{\tiny{JASA\_cal}}}$ & $\hat{\tau}_{\text{\tiny{Unadj}}}$ & $\hat{\tau}_{\text{\tiny{GComp}}}$ & $\hat{\tau}_{\text{\tiny{Lasso}}}$ & $\hat{\tau}_{\text{\tiny{CF}}}$ & $\hat{\tau}_{\text{\tiny{CF\_Lasso}}}$ & $\hat{\tau}^\dagger_{\text{\tiny{HOIF}}}$ & $\hat{\tau}^\ddagger_{\text{\tiny{HOIF}}}$ & $\hat{\tau}_{\text{\tiny{JASA}}}$ & $\hat{\tau}_{\text{\tiny{JASA\_cal}}}$ \\
& \multicolumn{9}{c}{\textbf{Bias}} & \multicolumn{9}{c}{\textbf{SD}} \\
\cmidrule(lr){2-10} \cmidrule(lr){11-19}
\multicolumn{19}{c}{$n$=50} \\
0.05 & 0.015 & -0.008 & -0.008 & -0.007 & -0.007 & 0.014 & -0.005 & 0.027 & 0.042 & 0.642 & 0.380 & 0.380 & 0.380 & 0.380 & 0.401 & 0.395 & 0.351 & 0.369 \\
  0.1 & -0.003 & 0.001 & 0.001 & 0.001 & 0.002 & 0.013 & -0.004 & 0.015 & 0.040 & 0.674 & 0.394 & 0.394 & 0.395 & 0.396 & 0.411 & 0.407 & 0.416 & 0.423 \\
  0.2 & -0.026 & -0.023 & -0.024 & -0.023 & -0.024 & -0.001 & -0.021 & 0.022 & 0.040 & 0.732 & 0.490 & 0.495 & 0.491 & 0.496 & 0.511 & 0.501 & 0.505 & 0.515 \\
  0.3 & 0.001 & -0.002 & -0.004 & -0.001 & -0.005 & 0.012 & -0.001 & 0.031 & 0.041 & 0.788 & 0.540 & 0.548 & 0.544 & 0.548 & 0.558 & 0.557 & 0.590 & 0.590 \\
  0.4 & 0.035 & 0.020 & 0.005 & 0.017 & 0.013 & 0.037 & 0.029 & 0.030 & 0.034 & 0.838 & 0.625 & 0.634 & 0.633 & 0.623 & 0.637 & 0.632 & 0.631 & 0.613 \\
  0.7 & 0.025 & 0.022 & 0.049 & 0.074 & 0.030 & 0.031 & 0.029 & 0.024 & 0.002 & 0.935 & 1.043 & 0.853 & 1.369 & 0.805 & 0.822 & 0.815 & 0.782 & 0.783 \\
\multicolumn{19}{c}{$n$=100} \\
0.05 & 0.001 & 0.005 & 0.005 & 0.005 & 0.005 & 0.015 & 0.004 & -0.009 & 0.006 & 0.432 & 0.278 & 0.279 & 0.278 & 0.278 & 0.283 & 0.278 & 0.285 & 0.290 \\
  0.1 & 0.011 & 0.017 & 0.017 & 0.017 & 0.017 & 0.023 & 0.016 & 0.010 & 0.028 & 0.521 & 0.312 & 0.312 & 0.312 & 0.312 & 0.320 & 0.317 & 0.327 & 0.330 \\
  0.2 & -0.005 & -0.009 & -0.009 & -0.010 & -0.011 & 0.004 & -0.007 & 0.012 & 0.024 & 0.561 & 0.377 & 0.377 & 0.378 & 0.379 & 0.393 & 0.383 & 0.380 & 0.377 \\
  0.3 & 0.000 & -0.010 & -0.009 & -0.008 & -0.008 & 0.001 & -0.008 & 0.008 & 0.015 & 0.601 & 0.415 & 0.416 & 0.418 & 0.421 & 0.416 & 0.411 & 0.444 & 0.446 \\
  0.4 & -0.022 & -0.007 & -0.004 & -0.014 & -0.013 & -0.014 & -0.017 & -0.004 & -0.009 & 0.640 & 0.472 & 0.481 & 0.477 & 0.469 & 0.465 & 0.465 & 0.477 & 0.470 \\
  0.7 & -0.027 & -0.031 & -0.023 & -0.062 & -0.022 & -0.019 & -0.025 & 0.044 & 0.026 & 0.717 & 0.793 & 0.667 & 0.936 & 0.617 & 0.638 & 0.630 & 0.610 & 0.608 \\
\multicolumn{19}{c}{$n$=200} \\
0.05 & 0.021 & 0.003 & 0.003 & 0.003 & 0.003 & 0.009 & 0.004 & -0.000 & 0.009 & 0.354 & 0.217 & 0.217 & 0.217 & 0.217 & 0.221 & 0.218 & 0.219 & 0.221 \\
  0.1 & -0.006 & -0.011 & -0.011 & -0.011 & -0.011 & -0.007 & -0.011 & 0.003 & 0.013 & 0.414 & 0.249 & 0.249 & 0.248 & 0.249 & 0.254 & 0.251 & 0.257 & 0.257 \\
  0.2 & -0.018 & -0.006 & -0.006 & -0.007 & -0.008 & -0.004 & -0.008 & 0.007 & 0.013 & 0.458 & 0.298 & 0.298 & 0.300 & 0.300 & 0.302 & 0.299 & 0.308 & 0.309 \\
  0.3 & -0.014 & -0.010 & -0.010 & -0.010 & -0.010 & -0.010 & -0.012 & 0.010 & 0.015 & 0.488 & 0.340 & 0.340 & 0.343 & 0.344 & 0.335 & 0.332 & 0.341 & 0.340 \\
  0.4 & -0.048 & -0.015 & -0.014 & -0.017 & -0.017 & -0.027 & -0.029 & 0.015 & 0.015 & 0.513 & 0.391 & 0.394 & 0.391 & 0.390 & 0.379 & 0.376 & 0.396 & 0.389 \\
  0.7 & -0.010 & -0.010 & -0.001 & -0.008 & 0.010 & -0.007 & -0.010 & -0.013 & -0.024 & 0.595 & 0.620 & 0.542 & 0.712 & 0.508 & 0.520 & 0.516 & 0.508 & 0.498\\
\midrule
& \multicolumn{9}{c}{\textbf{SE}} & \multicolumn{9}{c}{\textbf{COV (\%)}} \\
\cmidrule(lr){2-10} \cmidrule(lr){11-19}
\multicolumn{19}{c}{$n$=50} \\
0.05 & 0.609 & 0.337 & 0.337 & 0.372 & 0.372 & 0.375 & 0.381 & 0.342 & 0.350 & 93.3 & 90.7 & 90.7 & 93.4 & 93.4 & 93.0 & 94.1 & 93.9 & 93.3 \\
  0.1 & 0.658 & 0.356 & 0.356 & 0.414 & 0.414 & 0.418 & 0.421 & 0.415 & 0.409 & 93.5 & 92.2 & 92.1 & 95.8 & 95.7 & 95.6 & 95.6 & 94.4 & 93.5 \\
  0.2 & 0.734 & 0.374 & 0.374 & 0.504 & 0.509 & 0.497 & 0.498 & 0.495 & 0.478 & 95.5 & 85.9 & 85.3 & 94.8 & 94.7 & 94.2 & 94.6 & 93.9 & 91.8 \\
  0.3 & 0.782 & 0.370 & 0.373 & 0.591 & 0.602 & 0.563 & 0.562 & 0.557 & 0.533 & 93.7 & 81.4 & 81.0 & 95.8 & 95.6 & 94.8 & 95.2 & 93.3 & 92.9 \\
  0.4 & 0.814 & 0.357 & 0.371 & 0.703 & 0.702 & 0.623 & 0.620 & 0.625 & 0.590 & 92.9 & 71.1 & 71.5 & 97.1 & 96.5 & 93.4 & 93.6 & 94.1 & 94.0 \\
  0.7 & 0.894 & 0.264 & 0.373 & 2.150 & 0.927 & 0.800 & 0.788 & 0.799 & 0.739 & 93.5 & 38.1 & 60.2 & 99.7 & 96.5 & 93.9 & 94.0 & 95.2 & 93.4 \\
\multicolumn{19}{c}{$n$=100} \\
0.05 & 0.465 & 0.261 & 0.261 & 0.280 & 0.280 & 0.280 & 0.281 & 0.281 & 0.280 & 96.7 & 93.7 & 93.6 & 95.2 & 95.2 & 94.9 & 95.3 & 93.4 & 93.3 \\
  0.1 & 0.516 & 0.282 & 0.282 & 0.322 & 0.322 & 0.321 & 0.321 & 0.323 & 0.319 & 93.4 & 91.1 & 91.1 & 95.4 & 95.4 & 95.2 & 95.3 & 94.8 & 94.0 \\
  0.2 & 0.575 & 0.294 & 0.294 & 0.388 & 0.390 & 0.379 & 0.379 & 0.381 & 0.374 & 95.0 & 86.8 & 86.8 & 95.1 & 95.1 & 94.4 & 95.2 & 94.6 & 94.3 \\
  0.3 & 0.615 & 0.294 & 0.295 & 0.458 & 0.464 & 0.432 & 0.430 & 0.430 & 0.417 & 95.7 & 84.6 & 84.6 & 96.0 & 95.7 & 95.1 & 95.6 & 94.7 & 93.2 \\
  0.4 & 0.646 & 0.287 & 0.294 & 0.544 & 0.542 & 0.484 & 0.481 & 0.481 & 0.463 & 95.4 & 77.9 & 77.4 & 97.2 & 96.8 & 96.3 & 95.5 & 94.6 & 94.0 \\
  0.7 & 0.716 & 0.218 & 0.291 & 1.358 & 0.706 & 0.625 & 0.619 & 0.623 & 0.592 & 95.4 & 41.3 & 61.1 & 99.3 & 96.6 & 94.4 & 94.6 & 95.1 & 93.7 \\
\multicolumn{19}{c}{$n$=200} \\
0.05 & 0.368 & 0.207 & 0.207 & 0.221 & 0.221 & 0.220 & 0.220 & 0.219 & 0.218 & 96.0 & 93.8 & 93.8 & 95.4 & 95.4 & 95.5 & 95.8 & 94.5 & 94.1 \\
  0.1 & 0.408 & 0.223 & 0.223 & 0.253 & 0.253 & 0.251 & 0.251 & 0.251 & 0.250 & 95.4 & 91.8 & 91.8 & 96.2 & 96.2 & 94.7 & 94.7 & 94.4 & 94.2 \\
  0.2 & 0.458 & 0.235 & 0.235 & 0.307 & 0.307 & 0.298 & 0.297 & 0.298 & 0.294 & 95.0 & 87.8 & 87.7 & 95.4 & 95.4 & 95.4 & 95.5 & 94.1 & 93.6 \\
  0.3 & 0.490 & 0.235 & 0.235 & 0.359 & 0.362 & 0.338 & 0.337 & 0.340 & 0.333 & 94.7 & 83.3 & 83.8 & 95.5 & 95.2 & 94.8 & 94.5 & 95.3 & 94.4 \\
  0.4 & 0.517 & 0.229 & 0.232 & 0.428 & 0.425 & 0.379 & 0.377 & 0.380 & 0.370 & 94.9 & 73.3 & 73.7 & 96.9 & 96.8 & 94.8 & 95.0 & 94.1 & 94.4 \\
  0.7 & 0.576 & 0.180 & 0.232 & 1.004 & 0.567 & 0.500 & 0.497 & 0.502 & 0.483 & 94.7 & 42.1 & 59.8 & 99.5 & 97.1 & 95.4 & 94.9 & 94.5 & 93.2 \\
                \bottomrule
            \end{tabular}
        }
    \end{center}
    \label{table:cont_s1_bias_SD_SE_COV}
\end{table}

\newpage

\begin{table}[h!]
    \caption{Estimated bias, Monte Carlo standard deviation, average standard errors, and coverage probabilities for ATE obtained from Setting 3 with continuous outcomes.}
    \begin{center}
\resizebox{\textwidth}{!}{
\begin{tabular}{ccccccccccccccccccc}
\toprule
$k$ & $\hat{\tau}_{\text{\tiny{Unadj}}}$ & $\hat{\tau}_{\text{\tiny{GComp}}}$ & $\hat{\tau}_{\text{\tiny{Lasso}}}$ & $\hat{\tau}_{\text{\tiny{CF}}}$ & $\hat{\tau}_{\text{\tiny{CF\_Lasso}}}$ & $\hat{\tau}^\dagger_{\text{\tiny{HOIF}}}$ & $\hat{\tau}^\ddagger_{\text{\tiny{HOIF}}}$ & $\hat{\tau}_{\text{\tiny{JASA}}}$ & $\hat{\tau}_{\text{\tiny{JASA\_cal}}}$ & $\hat{\tau}_{\text{\tiny{Unadj}}}$ & $\hat{\tau}_{\text{\tiny{GComp}}}$ & $\hat{\tau}_{\text{\tiny{Lasso}}}$ & $\hat{\tau}_{\text{\tiny{CF}}}$ & $\hat{\tau}_{\text{\tiny{CF\_Lasso}}}$ & $\hat{\tau}^\dagger_{\text{\tiny{HOIF}}}$ & $\hat{\tau}^\ddagger_{\text{\tiny{HOIF}}}$ & $\hat{\tau}_{\text{\tiny{JASA}}}$ & $\hat{\tau}_{\text{\tiny{JASA\_cal}}}$ \\
& \multicolumn{9}{c}{\textbf{Bias}} & \multicolumn{9}{c}{\textbf{SD}} \\
\cmidrule(lr){2-10} \cmidrule(lr){11-19}
0.05 & 0.000 & 0.001 & 0.002 & 0.002 & 0.002 & 0.047 & 0.002 & -0.045 & -0.030 & 0.171 & 0.088 & 0.089 & 0.088 & 0.089 & 1.505 & 0.090 & 1.553 & 0.372 \\
  0.1 & -0.044 & -0.022 & -0.027 & -0.027 & -0.024 & 0.202 & -0.030 & 0.064 & 0.008 & 1.705 & 0.881 & 0.875 & 0.876 & 0.871 & 2.731 & 0.952 & 2.787 & 1.579 \\
  0.2 & 0.030 & -0.007 & -0.005 & -0.005 & -0.006 & -0.124 & 0.009 & -0.042 & 0.008 & 1.701 & 0.942 & 0.916 & 0.942 & 0.908 & 3.916 & 1.020 & 3.880 & 1.698 \\
  0.3 & 0.159 & 0.015 & 0.012 & 0.021 & 0.025 & 0.162 & 0.038 & -0.033 & -0.015 & 2.042 & 1.205 & 1.155 & 1.191 & 1.141 & 4.098 & 1.331 & 4.275 & 2.033 \\
  0.4 & 0.049 & -0.067 & -0.053 & -0.056 & -0.010 & -0.192 & -0.035 & -0.061 & -0.007 & 3.128 & 2.085 & 1.958 & 2.072 & 1.915 & 5.227 & 2.241 & 5.056 & 3.130\\
  0.54 & -0.330 & 0.027 & -0.164 & -0.090 & -0.200 & -0.362 & -0.234 & -0.067 & -0.059 & 4.595 & 3.345 & 3.031 & 3.554 & 2.932 & 6.691 & 3.366 & 6.284 & 4.650 \\
\midrule
& \multicolumn{9}{c}{\textbf{SE}} & \multicolumn{9}{c}{\textbf{COV (\%)}} \\
\cmidrule(lr){2-10} \cmidrule(lr){11-19}
0.05 & 0.172 & 0.085 & 0.085 & 0.091 & 0.091 & 1.740 & 0.096 & 1.604 & 0.206 & 95.2 & 94.6 & 94.6 & 95.5 & 95.6 & 99.0 & 96.4 & 97.1 & 82.1 \\
  0.1 & 1.621 & 0.782 & 0.792 & 1.151 & 0.927 & 3.291 & 0.938 & 3.091 & 1.372 & 93.9 & 92.4 & 93.1 & 96.4 & 95.7 & 98.9 & 94.3 & 98.0 & 90.5 \\
  0.2 & 1.672 & 0.757 & 0.779 & 1.358 & 1.056 & 5.872 & 1.037 & 5.437 & 1.533 & 95.1 & 88.4 & 90.8 & 97.2 & 96.8 & 99.7 & 95.5 & 99.4 & 92.1 \\
  0.3 & 2.021 & 0.856 & 0.904 & 1.770 & 1.357 & 7.877 & 1.334 & 7.552 & 1.898 & 94.0 & 84.3 & 87.2 & 98.9 & 97.3 & 99.7 & 94.2 & 100.0 & 94.2 \\
  0.4 & 3.160 & 1.225 & 1.346 & 3.170 & 2.273 & 10.464 & 2.242 & 9.999 & 3.011 & 95.4 & 75.3 & 82.5 & 98.0 & 96.5 & 99.8 & 94.9 & 100.0 & 94.2 \\
  0.54 & 4.555 & 1.575 & 1.897 & 7.217 & 3.474 & 13.892 & 3.452 & 13.239 & 4.481 & 95.5 & 62.0 & 74.4 & 100.0 & 97.1 & 100.0 & 95.0 & 100.0 & 94.7 \\
\bottomrule
 \end{tabular}
        }
    \end{center}
    \label{table:cont_s3_bias_SD_SE_COV}
\end{table}

\clearpage
\subsubsection{Binary outcomes}
Results are provided in \autoref{table:bin_s1_bias_SD_SE_COV}, \autoref{table:appe_bin_s2_bias_SD_SE_COV_ch3} in Section \ref{appe_sim_results_ch3}, and \autoref{table:bin_s3_bias_SD_SE_COV} for Settings 1, 2, and 3, respectively. The G-computation estimator is less sensitive to bias compared to the MLE of the treatment-coefficient in a logistic regression model. This is because the G-computation estimator based on canonical GLM averages predicted means, which are calibrated to match the observed means in the sample (see, for example, \autoref{figure:dist_MLE_bin} and \autoref{figure:dist_G_Comp_bin}). Apart from this, conclusions are largely as before, with the variance estimator of the Firth-based G-computation estimator also becoming invalid as the number of covariates increases, as a result of own-observation bias. The results reconfirm the importance of cross-fitting. However, when $n = 50$, coverage probabilities are still below the nominal level when using cross-fitting with or without post-LASSO. This may be due to the small sample size, which can lead to poor performance in estimating the fluctuation parameters within the TMLE procedure \citep{smith2025performance}. At $n$=200, their performance improves substantially (see \autoref{table:bin_s1_bias_SD_SE_COV} and \autoref{table:appe_bin_s2_bias_SD_SE_COV_ch3} in Section \ref{appe_sim_results_ch3}). As before, HOIF-based estimators (still based on linear regression within each treatment group) with centered outcome perform best, but are competitive with both JASA estimators; here, the calibrated JASA estimator yields coverage probability slightly below the nominal level. Results for Setting 3 are comparable to those for Setting 1 and Setting 2 despite the lower event rates (with success rate under treatment ranging form 6\% to 29\% (see the second column of \autoref{table:bin_s3_bias_SD_SE_COV}).

\begin{table}[!h]
    \caption{Estimated bias, Monte Carlo standard deviation, average standard errors, and coverage probabilities for ATE obtained from Setting 1 with binary outcomes.}
    \begin{center}
        \resizebox{\textwidth}{!}{
            \begin{tabular}{ccccccccccccccccccccccc}
                \toprule
$k$ & $\E\left(Y^1\right)$ & $\E\left(Y^0\right)$ & $\hat{\tau}_{\text{\tiny{Unadj}}}$ & $\hat{\tau}_{\text{\tiny{GComp}}}$ & $\hat{\tau}_{\text{\tiny{Firth}}}$ & $\hat{\tau}_{\text{\tiny{Lasso}}}$ & $\hat{\tau}_{\text{\tiny{CF}}}$ & $\hat{\tau}_{\text{\tiny{CF\_Lasso}}}$ & $\hat{\tau}^\dagger_{\text{\tiny{HOIF}}}$ & $\hat{\tau}^\ddagger_{\text{\tiny{HOIF}}}$ & $\hat{\tau}_{\text{\tiny{JASA}}}$ & $\hat{\tau}_{\text{\tiny{JASA\_cal}}}$ & $\hat{\tau}_{\text{\tiny{Unadj}}}$ & $\hat{\tau}_{\text{\tiny{GComp}}}$ & $\hat{\tau}_{\text{\tiny{Firth}}}$ & $\hat{\tau}_{\text{\tiny{Lasso}}}$ & $\hat{\tau}_{\text{\tiny{CF}}}$ & $\hat{\tau}_{\text{\tiny{CF\_Lasso}}}$ & $\hat{\tau}^\dagger_{\text{\tiny{HOIF}}}$ & $\hat{\tau}^\ddagger_{\text{\tiny{HOIF}}}$ & $\hat{\tau}_{\text{\tiny{JASA}}}$ & $\hat{\tau}_{\text{\tiny{JASA\_cal}}}$ \\
& & & \multicolumn{10}{c}{\textbf{Bias}} & \multicolumn{10}{c}{\textbf{SD}} \\
\cmidrule(lr){4-13} \cmidrule(lr){14-23}
\multicolumn{23}{c}{$n$=50} \\
0.05 & 0.682 & 0.547 & -0.005 & -0.003 & 0.009 & -0.002 & 0.002 & 0.001 & 0.001 & -0.001 & 0.010 & 0.010 & 0.132 & 0.115 & 0.127 & 0.116 & 0.130 & 0.131 & 0.120 & 0.117 & 0.131 & 0.131 \\
  0.1 & 0.676 & 0.546 & 0.005 & 0.001 & 0.009 & 0.001 & 0.006 & 0.005 & 0.004 & 0.003 & 0.010 & 0.010 & 0.145 & 0.129 & 0.134 & 0.133 & 0.143 & 0.145 & 0.136 & 0.133 & 0.139 & 0.141 \\
  0.2 & 0.667 & 0.543 & -0.002 & -0.002 & -0.000 & -0.002 & 0.003 & 0.000 & 0.004 & 0.002 & -0.005 & -0.004 & 0.139 & 0.123 & 0.131 & 0.129 & 0.140 & 0.144 & 0.131 & 0.126 & 0.127 & 0.130 \\
  0.3 & 0.662 & 0.542 & -0.001 & -0.007 & 0.012 & -0.003 & 0.006 & 0.004 & 0.000 & -0.003 & 0.005 & 0.006 & 0.140 & 0.129 & 0.143 & 0.134 & 0.144 & 0.144 & 0.141 & 0.132 & 0.134 & 0.136 \\
  0.4 & 0.659 & 0.541 & -0.006 & -0.012 & 0.006 & -0.007 & -0.000 & -0.004 & -0.000 & -0.001 & 0.004 & 0.006 & 0.143 & 0.130 & 0.154 & 0.137 & 0.152 & 0.153 & 0.143 & 0.136 & 0.137 & 0.138 \\
  0.7 & 0.653 & 0.539 & -0.004 & -0.033 & -0.007 & -0.007 & 0.003 & 0.001 & -0.000 & -0.000 & -0.002 & -0.005 & 0.140 & 0.111 & 0.159 & 0.134 & 0.154 & 0.149 & 0.144 & 0.141 & 0.141 & 0.143 \\
\multicolumn{23}{c}{$n$=100} \\
0.05 & 0.675 & 0.545 & -0.001 & -0.003 & -0.001 & -0.003 & -0.002 & -0.002 & -0.001 & -0.002 & -0.000 & -0.001 & 0.098 & 0.085 & 0.089 & 0.085 & 0.088 & 0.089 & 0.086 & 0.086 & 0.091 & 0.090 \\
  0.1 & 0.667 & 0.543 & -0.001 & 0.000 & 0.006 & 0.001 & 0.002 & 0.001 & 0.002 & 0.000 & 0.004 & 0.005 & 0.096 & 0.085 & 0.089 & 0.085 & 0.088 & 0.089 & 0.087 & 0.085 & 0.093 & 0.092 \\
  0.2 & 0.659 & 0.541 & -0.001 & -0.005 & 0.005 & -0.005 & -0.003 & -0.004 & -0.001 & -0.002 & 0.002 & 0.004 & 0.096 & 0.090 & 0.093 & 0.091 & 0.095 & 0.095 & 0.093 & 0.089 & 0.093 & 0.094 \\
  0.3 & 0.655 & 0.540 & 0.003 & -0.002 & 0.005 & -0.001 & -0.000 & 0.001 & 0.001 & 0.001 & -0.004 & -0.004 & 0.097 & 0.092 & 0.106 & 0.092 & 0.099 & 0.097 & 0.093 & 0.090 & 0.091 & 0.092 \\
  0.4 & 0.652 & 0.539 & -0.002 & -0.007 & 0.004 & -0.006 & -0.004 & -0.002 & 0.001 & -0.001 & -0.006 & -0.005 & 0.101 & 0.096 & 0.114 & 0.094 & 0.102 & 0.102 & 0.097 & 0.093 & 0.095 & 0.095 \\
  0.7 & 0.647 & 0.538 & 0.002 & -0.037 & -0.000 & -0.005 & -0.001 & -0.000 & 0.000 & 0.000 & 0.002 & 0.002 & 0.098 & 0.076 & 0.112 & 0.097 & 0.104 & 0.103 & 0.099 & 0.098 & 0.097 & 0.096 \\
\multicolumn{23}{c}{$n$=200} \\
0.05 & 0.667 & 0.543 & 0.005 & 0.004 & 0.001 & 0.004 & 0.004 & 0.004 & 0.005 & 0.004 & 0.001 & 0.000 & 0.068 & 0.057 & 0.060 & 0.057 & 0.059 & 0.060 & 0.058 & 0.058 & 0.061 & 0.061 \\
  0.1 & 0.659 & 0.541 & -0.000 & -0.000 & -0.001 & -0.000 & 0.000 & 0.000 & 0.000 & -0.001 & -0.002 & -0.002 & 0.073 & 0.062 & 0.063 & 0.063 & 0.064 & 0.064 & 0.064 & 0.063 & 0.064 & 0.064 \\
  0.2 & 0.652 & 0.539 & -0.001 & -0.003 & 0.005 & -0.004 & -0.004 & -0.004 & -0.002 & -0.003 & 0.000 & 0.000 & 0.068 & 0.063 & 0.062 & 0.063 & 0.065 & 0.065 & 0.062 & 0.061 & 0.059 & 0.059 \\
  0.3 & 0.649 & 0.538 & -0.001 & 0.000 & 0.010 & -0.002 & -0.004 & -0.002 & -0.002 & -0.002 & 0.000 & 0.001 & 0.068 & 0.069 & 0.072 & 0.065 & 0.071 & 0.067 & 0.064 & 0.063 & 0.064 & 0.064 \\
  0.4 & 0.646 & 0.537 & -0.002 & -0.003 & 0.011 & -0.005 & -0.004 & -0.001 & -0.001 & -0.002 & -0.002 & -0.002 & 0.071 & 0.074 & 0.086 & 0.070 & 0.075 & 0.073 & 0.067 & 0.066 & 0.067 & 0.068 \\
  0.7 & 0.641 & 0.536 & 0.000 & -0.043 & -0.006 & -0.004 & 0.003 & 0.001 & 0.003 & 0.002 & -0.001 & -0.001 & 0.072 & 0.052 & 0.075 & 0.069 & 0.076 & 0.074 & 0.070 & 0.069 & 0.067 & 0.066 \\
\midrule
& & & \multicolumn{10}{c}{\textbf{SE}} & \multicolumn{10}{c}{\textbf{COV (\%)}} \\
\cmidrule(lr){4-13} \cmidrule(lr){14-23}
\multicolumn{23}{c}{$n$=50} \\
0.05 & 0.682 & 0.547 & 0.138 & 0.115 & 0.118 & 0.116 & 0.118 & 0.120 & 0.127 & 0.126 & 0.127 & 0.126 & 95.6 & 94.0 & 92.8 & 94.2 & 92.2 & 92.4 & 96.1 & 96.5 & 93.9 & 93.7 \\
  0.1 & 0.676 & 0.546 & 0.138 & 0.110 & 0.110 & 0.111 & 0.118 & 0.121 & 0.128 & 0.126 & 0.128 & 0.126 & 92.2 & 88.9 & 87.4 & 88.3 & 88.6 & 88.3 & 92.6 & 92.0 & 92.4 & 91.3 \\
  0.2 & 0.667 & 0.543 & 0.138 & 0.099 & 0.099 & 0.105 & 0.121 & 0.125 & 0.136 & 0.128 & 0.133 & 0.124 & 94.4 & 89.0 & 84.8 & 89.2 & 91.2 & 90.9 & 95.7 & 95.3 & 95.2 & 93.0 \\
  0.3 & 0.662 & 0.542 & 0.139 & 0.088 & 0.090 & 0.102 & 0.126 & 0.129 & 0.145 & 0.131 & 0.138 & 0.126 & 94.5 & 79.5 & 75.6 & 84.9 & 91.1 & 91.1 & 95.7 & 94.5 & 94.8 & 92.2 \\
  0.4 & 0.659 & 0.541 & 0.139 & 0.075 & 0.082 & 0.101 & 0.128 & 0.130 & 0.154 & 0.133 & 0.143 & 0.127 & 93.0 & 71.9 & 69.5 & 82.2 & 89.5 & 90.0 & 96.1 & 93.9 & 95.6 & 92.6 \\
  0.7 & 0.653 & 0.539 & 0.139 & 0.045 & 0.076 & 0.099 & 0.133 & 0.133 & 0.187 & 0.137 & 0.155 & 0.130 & 93.5 & 53.3 & 64.3 & 80.2 & 89.5 & 90.9 & 98.3 & 92.8 & 96.9 & 92.0 \\
\multicolumn{23}{c}{$n$=100} \\
0.05 & 0.675 & 0.545 & 0.097 & 0.080 & 0.080 & 0.080 & 0.083 & 0.084 & 0.086 & 0.086 & 0.087 & 0.086 & 93.5 & 92.3 & 91.0 & 92.0 & 92.4 & 92.8 & 94.1 & 94.2 & 93.5 & 93.2 \\
  0.1 & 0.667 & 0.543 & 0.097 & 0.076 & 0.076 & 0.077 & 0.084 & 0.086 & 0.088 & 0.087 & 0.089 & 0.086 & 94.8 & 92.1 & 91.1 & 91.9 & 94.1 & 94.1 & 94.9 & 95.6 & 93.3 & 92.0 \\
  0.2 & 0.659 & 0.541 & 0.098 & 0.068 & 0.068 & 0.071 & 0.088 & 0.090 & 0.092 & 0.088 & 0.092 & 0.088 & 94.6 & 85.2 & 85.4 & 86.8 & 92.9 & 93.1 & 94.8 & 94.6 & 94.6 & 93.1 \\
  0.3 & 0.655 & 0.540 & 0.098 & 0.059 & 0.061 & 0.068 & 0.093 & 0.093 & 0.099 & 0.090 & 0.096 & 0.090 & 94.9 & 78.5 & 72.9 & 83.9 & 93.2 & 92.9 & 95.8 & 94.7 & 96.7 & 94.7 \\
  0.4 & 0.652 & 0.539 & 0.098 & 0.049 & 0.055 & 0.067 & 0.096 & 0.095 & 0.106 & 0.092 & 0.100 & 0.091 & 94.6 & 67.3 & 63.9 & 80.8 & 92.2 & 92.8 & 96.9 & 94.2 & 96.0 & 93.3 \\
  0.7 & 0.647 & 0.538 & 0.098 & 0.025 & 0.052 & 0.065 & 0.100 & 0.098 & 0.129 & 0.096 & 0.108 & 0.093 & 93.8 & 38.7 & 62.2 & 75.7 & 93.6 & 93.4 & 99.2 & 94.1 & 97.6 & 94.1 \\
\multicolumn{23}{c}{$n$=200} \\
0.05 & 0.667 & 0.543 & 0.069 & 0.056 & 0.056 & 0.056 & 0.058 & 0.059 & 0.060 & 0.060 & 0.060 & 0.059 & 94.8 & 93.6 & 92.7 & 93.5 & 94.4 & 94.3 & 95.7 & 95.5 & 94.8 & 94.2 \\
  0.1 & 0.659 & 0.541 & 0.069 & 0.052 & 0.053 & 0.053 & 0.059 & 0.060 & 0.061 & 0.060 & 0.061 & 0.060 & 92.8 & 90.0 & 89.3 & 90.2 & 93.4 & 93.2 & 93.6 & 93.1 & 92.9 & 92.4 \\
  0.2 & 0.652 & 0.539 & 0.069 & 0.047 & 0.047 & 0.048 & 0.063 & 0.064 & 0.064 & 0.061 & 0.064 & 0.062 & 94.9 & 84.3 & 86.1 & 85.8 & 93.4 & 94.2 & 94.9 & 94.6 & 96.9 & 96.6 \\
  0.3 & 0.649 & 0.538 & 0.069 & 0.040 & 0.042 & 0.046 & 0.068 & 0.066 & 0.068 & 0.063 & 0.067 & 0.063 & 95.2 & 74.7 & 73.3 & 82.2 & 94.3 & 94.6 & 96.5 & 95.5 & 95.8 & 94.4 \\
  0.4 & 0.646 & 0.537 & 0.069 & 0.032 & 0.037 & 0.044 & 0.071 & 0.068 & 0.073 & 0.064 & 0.070 & 0.064 & 93.6 & 59.6 & 60.1 & 77.0 & 92.7 & 91.7 & 95.7 & 93.7 & 95.6 & 93.5 \\
  0.7 & 0.641 & 0.536 & 0.069 & 0.013 & 0.036 & 0.044 & 0.073 & 0.071 & 0.090 & 0.067 & 0.076 & 0.066 & 94.0 & 25.0 & 64.7 & 73.7 & 94.1 & 93.1 & 98.4 & 93.2 & 97.1 & 94.8 \\
                \bottomrule
            \end{tabular}
        }
    \end{center}
    \label{table:bin_s1_bias_SD_SE_COV}
\end{table}

\newpage
\begin{table}[!h]
    \caption{Estimated bias, Monte Carlo standard deviation, average standard errors, and coverage probabilities for ATE obtained from Setting 3 with binary outcomes.}
    \begin{center}
        \resizebox{\textwidth}{!}{
            \begin{tabular}{ccccccccccccccccccccccc}
                \toprule
$k$ & $\E\left(Y^1\right)$ & $\E\left(Y^0\right)$ & $\hat{\tau}_{\text{\tiny{Unadj}}}$ & $\hat{\tau}_{\text{\tiny{GComp}}}$ & $\hat{\tau}_{\text{\tiny{Firth}}}$ & $\hat{\tau}_{\text{\tiny{Lasso}}}$ & $\hat{\tau}_{\text{\tiny{CF}}}$ & $\hat{\tau}_{\text{\tiny{CF\_Lasso}}}$ & $\hat{\tau}^\dagger_{\text{\tiny{HOIF}}}$ & $\hat{\tau}^\ddagger_{\text{\tiny{HOIF}}}$ & $\hat{\tau}_{\text{\tiny{JASA}}}$ & $\hat{\tau}_{\text{\tiny{JASA\_cal}}}$ & $\hat{\tau}_{\text{\tiny{Unadj}}}$ & $\hat{\tau}_{\text{\tiny{GComp}}}$ & $\hat{\tau}_{\text{\tiny{Firth}}}$ & $\hat{\tau}_{\text{\tiny{Lasso}}}$ & $\hat{\tau}_{\text{\tiny{CF}}}$ & $\hat{\tau}_{\text{\tiny{CF\_Lasso}}}$ & $\hat{\tau}^\dagger_{\text{\tiny{HOIF}}}$ & $\hat{\tau}^\ddagger_{\text{\tiny{HOIF}}}$ & $\hat{\tau}_{\text{\tiny{JASA}}}$ & $\hat{\tau}_{\text{\tiny{JASA\_cal}}}$ \\
& & & \multicolumn{10}{c}{\textbf{Bias}} & \multicolumn{10}{c}{\textbf{SD}} \\
\cmidrule(lr){4-13} \cmidrule(lr){14-23}
0.05 & 0.060 & 0.153 & 0.001 & 0.002 & 0.010 & 0.001 & 0.001 & 0.001 & 0.000 & 0.001 & 0.010 & 0.010 & 0.054 & 0.054 & 0.052 & 0.054 & 0.056 & 0.056 & 0.055 & 0.054 & 0.052 & 0.052 \\
  0.1 & 0.062 & 0.157 & -0.000 & 0.001 & 0.009 & 0.000 & -0.001 & -0.001 & -0.001 & 0.000 & 0.010 & 0.012 & 0.058 & 0.058 & 0.054 & 0.058 & 0.061 & 0.060 & 0.058 & 0.058 & 0.054 & 0.055 \\
  0.2 & 0.293 & 0.362 & -0.000 & 0.006 & -0.046 & 0.005 & -0.004 & -0.003 & -0.000 & 0.001 & -0.043 & -0.042 & 0.084 & 0.050 & 0.068 & 0.050 & 0.065 & 0.061 & 0.064 & 0.062 & 0.069 & 0.069 \\
  0.3 & 0.250 & 0.329 & -0.000 & 0.010 & -0.038 & 0.006 & -0.001 & -0.000 & 0.000 & 0.000 & -0.032 & -0.033 & 0.084 & 0.055 & 0.071 & 0.057 & 0.066 & 0.065 & 0.066 & 0.064 & 0.068 & 0.069 \\
  0.4 & 0.266 & 0.343 & -0.001 & 0.010 & -0.020 & 0.004 & -0.003 & -0.003 & -0.000 & -0.000 & -0.013 & -0.013 & 0.084 & 0.056 & 0.076 & 0.058 & 0.068 & 0.066 & 0.067 & 0.065 & 0.070 & 0.070 \\
  0.54 & 0.286 & 0.356 & -0.000 & 0.012 & -0.004 & 0.007 & -0.001 & -0.000 & 0.001 & 0.001 & 0.005 & 0.005 & 0.085 & 0.054 & 0.077 & 0.059 & 0.073 & 0.074 & 0.071 & 0.070 & 0.071 & 0.071 \\
\midrule
& & & \multicolumn{10}{c}{\textbf{SE}} & \multicolumn{10}{c}{\textbf{COV (\%)}} \\
\cmidrule(lr){4-13} \cmidrule(lr){14-23}
0.05 & 0.060 & 0.153 & 0.054 & 0.051 & 0.049 & 0.052 & 0.053 & 0.053 & 0.055 & 0.055 & 0.053 & 0.052 & 94.4 & 93.5 & 92.7 & 93.9 & 93.0 & 93.3 & 94.9 & 95.1 & 94.0 & 93.7 \\
  0.1 & 0.062 & 0.157 & 0.055 & 0.050 & 0.048 & 0.052 & 0.055 & 0.055 & 0.056 & 0.056 & 0.055 & 0.052 & 94.2 & 91.0 & 90.8 & 92.8 & 91.8 & 92.8 & 94.8 & 94.9 & 94.2 & 93.8 \\
  0.2 & 0.293 & 0.362 & 0.084 & 0.035 & 0.051 & 0.039 & 0.052 & 0.051 & 0.064 & 0.062 & 0.068 & 0.065 & 94.6 & 81.9 & 75.6 & 85.5 & 87.8 & 88.9 & 95.2 & 95.1 & 90.6 & 89.3 \\
  0.3 & 0.250 & 0.329 & 0.081 & 0.035 & 0.048 & 0.042 & 0.058 & 0.057 & 0.067 & 0.064 & 0.069 & 0.067 & 93.6 & 76.0 & 76.8 & 83.4 & 91.3 & 92.0 & 94.7 & 94.7 & 92.5 & 91.6 \\
  0.4 & 0.266 & 0.343 & 0.083 & 0.033 & 0.046 & 0.040 & 0.060 & 0.060 & 0.071 & 0.066 & 0.070 & 0.068 & 95.7 & 72.4 & 75.0 & 80.6 & 91.5 & 92.6 & 96.5 & 95.3 & 95.0 & 95.3 \\
  0.54 & 0.286 & 0.356 & 0.084 & 0.031 & 0.047 & 0.040 & 0.065 & 0.066 & 0.077 & 0.069 & 0.069 & 0.067 & 94.3 & 72.3 & 74.3 & 79.1 & 91.6 & 90.8 & 95.9 & 93.8 & 94.9 & 93.8 \\
                \bottomrule
            \end{tabular}
        }
    \end{center}
    \label{table:bin_s3_bias_SD_SE_COV}
\end{table}

In terms of power, GComp, post-Lasso and Firth show the highest power, especially for larger $k$; however, since these estimators inflate the type I error, this higher power is partly spurious. The CF, HOIF and JASA-type estimators have lower power than these, particularly as k increases, but they remain more powerful than the unadjusted estimator (see \autoref{table:appe_cont_s1_power_ch3} and \autoref{table:appe_bin_s1_power_ch3}).

Overall, the simulation studies show that cross-fitting combined with post-LASSO G-computation, HOIF-based estimators with centered outcomes, and calibrated JASA-based estimators demonstrate reliable performance across all considered settings for both continuous and binary outcomes. For continuous outcomes, the sandwich variance estimators, particularly HC1-type, also perform well when a linear outcome model with main effects of covariates and no treatment-by-covariate interactions is used (see Section \ref{appe_sandwich_variance_ch3} for the procedure and \autoref{table:appe_cont_s1_WCI_HC_SE_corr_COV_ch3} in Section \ref{appe_sim_results_ch3} for results).

\clearpage
\section{Data analysis}\label{sec:data_analysis_ch3}
\subsection{Description of the data}
We analyze data from the Best Apnea Interventions for Research (BestAIR) trial \citep{zhao2017effect} to illustrate the performance of the estimators under consideration. The trial randomized 169 patients diagnosed with sleep apnea who either had established cardiovascular disease (CVD) or presented with at least three CVD risk factors, into four arms. The two active treatment arms are: 1) active CPAP (Continuous Positive Airway Pressure) therapy supported by a trained sleep technician, and 2) active CPAP combined with a behavioral intervention delivered by a behavioral therapist. Further, the two control arms are: 1) sham-CPAP, and 2) conservative medical therapy (CMT) alone. We consider the combined treatment and the combined control arms, as in the original study \citep{zhao2017effect}, and later used in \citep{liu2026coadvise}.

The primary endpoint is the change in mean 24-hour systolic blood pressure (SBP). In our analysis, we consider two outcome measures: (i) continuous SBP values measured in mmHg at six months, and (ii) binary SBP outcomes at six months, categorizing SBP as high ($\geq$130 mmHg) or low ($<$130 mmHg). The data set is high-dimensional, as it consists of 114 baseline covariates for a sample size of 169. After excluding observations with missing outcomes, the final analysis consists of 124 individuals. Therefore, all analyses reported below are based on this complete-case sample. We further exclude covariates with 10 or more missing values. We finally rely on 67 baseline covariates. Missing covariate values are imputed using mean imputation. This does not induce bias or excess variability as long as the imputation process is carried out without incorporating treatment or outcome information, the imputed covariates remain independent of the treatment assignment \citep{van2024covariate,zhao2024adjust}.

\subsection{Results}
We consider all ATE estimators and their corresponding variance estimators explained in Section \ref{subsec:estimators_and_performance_measures_ch3}. To compare the performance of estimators across different covariate-to-sample size ratios, we considered three covariate sets of sizes 6, 50, and 67. These subsets are selected from the full set of 67 baseline covariates using the \texttt{bestsubset} R package according to mean squared error. These choices correspond to covariate-to-sample size ratios of $k=0.05$, $0.40$, and $0.54$, respectively.

\autoref{table:DA} presents the results of the effect of CPAP on continuous and binary SBP. The effect estimate and confidence interval of the unadjusted estimator remains the same across subsets, since it does not include covariates. The standard G-computation estimators give narrowest confidence intervals with both continuous and binary SBP at all values of $k$. However, it may underestimate the true variability of the estimated treatment effect as shown in the simulations. The Firth-based G-computation estimators and post-LASSO G-computation estimators exhibit confidence interval widths comparable to those of the standard G-computation estimators, but may underestimate the true variability of the estimated treatment effect as demonstrated in the simulations.
\begin{table}[!h]
	\caption{Average treatment effect estimates, 95\% confidence intervals and width of confidence intervals for the effect of CPAP on 24-hour systolic blood pressure (SBP) in the BestAIR study. }
	\begin{center}
			\begin{tabular}{ccccccc}
				\toprule
            &\multicolumn{3}{c}{\textbf{Continuous SBP}} & \multicolumn{3}{c}{\textbf{Binary SBP}}  \\
				\cmidrule(lr){2-4}  \cmidrule(lr){5-7}
				Estimator & Estimate & CIs & WCI & Estimate & CIs & WCI\\
				\midrule
\multicolumn{7}{c}{$k$=0.05} \\
$\hat{\tau}_{\text{\tiny{Unadj}}}$ &  -4.91 & (-9.35, -0.47) & 8.88  & -0.21 & (-0.37, -0.05) & 0.32\\ 
  $\hat{\tau}_{\text{\tiny{GComp}}}$ & -1.99 & (-4.89, 0.91) & 5.79 & -0.10 & (-0.21, 0.01) & 0.22 \\ 
  $\hat{\tau}_{\text{\tiny{Firth}}}$ &--&--&--&-0.11 & (-0.22, 0.00) & 0.22 \\ 
  $\hat{\tau}_{\text{\tiny{Lasso}}}$ & -2.00 & (-4.89, 0.90) & 5.79 & -0.10 & (-0.21, 0.01) & 0.22 \\ 
  $\hat{\tau}_{\text{\tiny{CF}}}$ & -2.00 & (-5.01, 1.02) & 6.02 & -0.09 & (-0.21, 0.03) & 0.24\\ 
  $\hat{\tau}_{\text{\tiny{CF\_Lasso}}}$ & -2.00 & (-5.01, 1.02) & 6.02 & -0.10 & (-0.22, 0.02) & 0.24\\ 
  $\hat{\tau}^\dagger_{\text{\tiny{HOIF}}}$ &  -3.08 & (-7.36, 1.20) & 8.56 & -0.15 & (-0.27, -0.02) & 0.25 \\ 
  $\hat{\tau}^\ddagger_{\text{\tiny{HOIF}}}$ & -2.11 & (-5.30, 1.08) & 6.38 & -0.15 & (-0.28, -0.01) & 0.27\\ 
  $\hat{\tau}_{\text{\tiny{JASA}}}$ & -3.07 & (-7.16, 1.03) & 8.19 & -0.14 & (-0.27, -0.01) & 0.26 \\ 
  $\hat{\tau}_{\text{\tiny{JASA\_cal}}}$ & -3.64 & (-7.00, -0.28) & 6.71 & -0.12 & (-0.26, 0.01) & 0.27 \\ 
\multicolumn{7}{c}{$k$=0.4} \\
$\hat{\tau}_{\text{\tiny{Unadj}}}$ & -4.91 & (-9.35, -0.47) & 8.88 & -0.21 & (-0.37, -0.05) & 0.32\\ 
  $\hat{\tau}_{\text{\tiny{GComp}}}$ & -1.91 & (-4.17, 0.35) & 4.52 & -0.06 & (-0.11, 0.00) & 0.11  \\ 
 $\hat{\tau}_{\text{\tiny{Firth}}}$ &--&--&--&-0.11 & (-0.19, -0.04) & 0.15 \\ 
  $\hat{\tau}_{\text{\tiny{Lasso}}}$ & -4.17 & (-7.18, -1.16) & 6.02 & -0.14 & (-0.27, -0.01) & 0.26 \\ 
  $\hat{\tau}_{\text{\tiny{CF}}}$ & -2.25 & (-8.93, 4.43) & 13.36 & -0.09 & (-0.25, 0.08) & 0.33 \\ 
  $\hat{\tau}_{\text{\tiny{CF\_Lasso}}}$ & -3.93 & (-7.65, -0.22) & 7.43 & -0.12 & (-0.27, 0.02) & 0.29 \\ 
  $\hat{\tau}^\dagger_{\text{\tiny{HOIF}}}$ & -0.71 & (-20.62, 19.21) & 39.83 & -0.14 & (-0.29, 0.01) & 0.30 \\ 
  $\hat{\tau}^\ddagger_{\text{\tiny{HOIF}}}$ & -3.51 & (-7.17, 0.16) & 7.32 &  -0.15 & (-0.29, -0.01) & 0.28  \\ 
 $\hat{\tau}_{\text{\tiny{JASA}}}$ & -0.65 & (-20.09, 18.8) & 38.89 &-0.15 & (-0.31, 0.00) & 0.31 \\ 
  $\hat{\tau}_{\text{\tiny{JASA\_cal}}}$ & -4.16 & (-8.54, 0.22) & 8.76 & -0.18 & (-0.33, -0.03) & 0.30 \\ 
\multicolumn{7}{c}{$k$=0.54} \\
$\hat{\tau}_{\text{\tiny{Unadj}}}$ & -4.91 & (-9.35, -0.47) & 8.88 & -0.21 & (-0.37, -0.05) & 0.32\\ 
  $\hat{\tau}_{\text{\tiny{GComp}}}$ & -2.45 & (-4.70, -0.20) & 4.50  & -0.07 & (-0.12, -0.02) & 0.10 \\ 
 $\hat{\tau}_{\text{\tiny{Firth}}}$ &--&--&--&  -0.17 & (-0.25, -0.08) & 0.17 \\ 
  $\hat{\tau}_{\text{\tiny{Lasso}}}$ & -3.61 & (-6.94, -0.28) & 6.65  & -0.16 & (-0.26, -0.06) & 0.20 \\ 
  $\hat{\tau}_{\text{\tiny{CF}}}$ & -2.23 & (-9.15, 4.70) & 13.86  & -0.10 & (-0.25, 0.06) & 0.32 \\ 
  $\hat{\tau}_{\text{\tiny{CF\_Lasso}}}$ & -3.94 & (-7.77, -0.11) & 7.66   &  -0.15 & (-0.30, 0.00) & 0.30\\ 
  $\hat{\tau}^\dagger_{\text{\tiny{HOIF}}}$ &  -1.01 & (-26.49, 24.47) & 50.96  & -0.17 & (-0.34, 0.00) & 0.34 \\ 
  $\hat{\tau}^\ddagger_{\text{\tiny{HOIF}}}$ & -4.04 & (-8.19, 0.12) & 8.31 & -0.19 & (-0.35, -0.04) & 0.30 \\ 
   $\hat{\tau}_{\text{\tiny{JASA}}}$ & -0.62 & (-25.63, 24.38) & 50.01 & -0.16 & (-0.32, 0.00) & 0.32  \\ 
  $\hat{\tau}_{\text{\tiny{JASA\_cal}}}$ & -4.46 & (-8.87, -0.06) & 8.81 & -0.19 & (-0.34, -0.04) & 0.31 \\
				\bottomrule						
			\end{tabular}
	\end{center}
	\label{table:DA}
\end{table}

Cross-fitting G-computation (without lasso), HOIF-based estimators with uncentered outcome and uncalibrated JASA-estimators provide wide confidence intervals, showing instability at $k$=0.4 and 0.54. Cross-fitting combined with the post-LASSO G-computation, the HOIF-based estimators with centered outcomes  and the calibrated JASA-estimators yield confidence intervals narrower than unadjusted estimators at all values of $k$ with both endpoints. This finding is consistent with the simulation results, suggesting that these estimators may better capture the variability of the estimated ATE (as seen in simulations).

Overall, smaller subsets (i.e., smaller $k$) yield narrower intervals, which may be because the most prognostic covariates are already accounted for at $k=0.05$. We also note that the treatment effect estimates vary with the value of $k$ across estimators (see \autoref{table:DA}). This variation could be explained by the imprecision and covariate imbalance between treatment arms in the BestAIR trial, as discussed by \cite{gao2024does}.

\section{Discussion}\label{sec:discussion_ch3}
In this paper, we provided accessible insights into methods for adjusting for a large number of covariates relative to the sample size in RCTs and assessed their performance through independent numerical experiments. We first examined standard G-computation estimators for estimating the ATE, which are known to be robust to model misspecification when implemented using canonical GLMs \citep{FDA2023,van2024covariate}. We then explored simple and more recent advanced methods for mitigating own-observation bias arising from overfitting in the MLEs of GLMs.

G-computation estimators of the ATE are generally found to be less biased than the MLEs of GLM model parameters when the number of covariates is large relative to the sample size. This is because G-computation estimators average predicted means, which are calibrated to match the observed means, thereby reducing bias even in the presence of many covariates. However, their standard error estimators based on the influence function underestimates the true empirical variability of the estimated ATE when the number of covariates is large compared to the sample size and the same patients are used for both training the model and making outcome predictions. This occurs because these influence functions include the prediction errors, which are underestimated when predicting outcomes for the patients used for training the model \citep{cattaneo2019two}.

Although post-LASSO is known to reduce overfitting bias under sparsity assumptions, we found this to be insufficient at the low sample size that we considered. In contrast, cross-fitting combined with post-LASSO G-computation estimators performed well with small sample sizes and many covariates, consistent with \citep{van2024automated}. Notably, recently proposed leave-one-out cross-fitting based estimators \citep{zhao2024covariate, zhao2026adjusting} provide practical solutions for covariate adjustment in settings where the covariate dimension is large relative to the sample size. Among the methods considered, HOIF-based estimators with centered outcomes (and calibrated JASA) consistently performed best.

In simulations, we only considered correctly specified outcome models with continuous outcomes. While G-computation estimators with canonical GLMs are generally robust to model misspecification when applied to data from RCTs, their use under misspecification, especially in small sample sizes, can lead to reduced efficiency \citep{tackney2023comparison}. However, the HOIF-based estimators with binary outcomes were used under misspecified outcome models and nevertheless performed well in our simulations, suggesting robustness beyond the settings we formally evaluated.

For binary outcomes, the HOIF-based estimators (like AIPW estimators) may not remain within the same range of the true treatment effect, as their construction involves inverse probability weights and a hat matrix that are not inherently bounded. Considering TMLE to ensure the estimates remain within the same range of the true treatment effect might be of interest. Future work could also explore combining JASA and HOIF-based estimators with variable selection, which would likely improve efficiency in high-dimensional settings. Finally, extending covariate adjustment methods to accommodate high-dimensional covariates across a wider range of outcome types (e.g., for longitudinal outcomes) remains an important direction for future research.

\appendix
\renewcommand{\thesection}{\Alph{section}}
\renewcommand{\thesubsection}{\thesection.\arabic{subsection}}

\renewcommand{\thefigure}{\thesubsection.\arabic{figure}}
\renewcommand{\thetable}{\thesubsection.\arabic{table}}

\pretocmd{\subsection}{\setcounter{figure}{0}\setcounter{table}{0}}{}{}

\newpage	
\appendix
\renewcommand{\thesection}{\Alph{section}}
\renewcommand{\thesubsection}{\thesection.\arabic{subsection}}
		
\renewcommand{\thefigure}{\thesubsection.\arabic{figure}}
\renewcommand{\thetable}{\thesubsection.\arabic{table}}
		
\pretocmd{\subsection}{\setcounter{figure}{0}\setcounter{table}{0}}{}{}

\section{Theoretical derivations}
\subsection{G-computation estimators with many covariates}\label{appe_G_computation_with_many_covariates_ch3}
When canonical GLMs are used, G-computation estimators are numerically equivalent to AIPW estimators. Here, we utilize the AIPW representation for analytical convenience in our subsequent theoretical derivations. Consider the AIPW estimator of $\theta= \E\left(Y^1\right)$. The general theory on this estimator shows that, when sample splitting is used, it is root-$n$ consistent when the nuisance parameter estimators converge to some limit (in the sense that the outcome predictions $\hat{m}^1(X)$ of $\E\left(Y|A = 1, X\right)$ have a limit $m^1(X)$ such that $\E\left\{\hat{m}^1(X)-m^1(X)\right\}^2$ converges to zero) and that the second order remainder term 
\begin{align*}
   \E\left[\left(\frac{\pi}{\hat{\pi}}-1\right)\left\{\hat{m}^1(X)-m^1(X)\right\}\right]=o_{p}({n^{-1/2}}).
\end{align*}
Importantly, this limit may differ from $\E\left(Y|A = 1, X\right)$, though in that case, a slightly different standard error calculation may be needed \citep{van2024automated}. 
As we rely on settings where randomization probabilities are independent of baseline covariates, the remainder term equals
\begin{align*}
   \left(\frac{\pi}{\hat{\pi}}-1\right)\E\left\{\hat{m}^1(X)-m^1(X)\right\}\leq \left(\frac{\pi}{\hat{\pi}}-1\right)\E\left[\left\{\hat{m}^1(X)-m^1(X)\right\}^2\right],
\end{align*}
by the Cauchy-Schwarz inequality. Here, the first term is $O_{p}({n^{-1/2}})$. While the remaining requirement that $\E\left\{\hat{m}^1(X)-m^1(X)\right\}^2=o_p(1)$ is generally weak, this may not be so when the number of covariates is very large relative to the sample size.
For instance, MLEs in exponential family models are only consistent when $p/n$ converges to zero \citep{portnoy1988asymptotic}. In ordinary nonparametric regression, if the regression function has smoothness $s$, the minimax mean squared risk often behaves like
$n^{-2s/(2s+p)}$ \citep{stone1982optimal}. If now $p \asymp n$, then  $-2s/(2s+p)\to 0$, so the rate becomes roughly $o(1)$, which does not necessarily go to zero in any useful way.

The question remains whether this is just an artefact of the usual splitting of terms in the von Mises expansion into empirical process terms and remainder terms (in other words, it may be that their sum can be shown to be small, but not the individual components). Let us therefore proceed in more detail. Suppose we have a correctly specified outcome model so that $m^1(X)=\E\left(Y |A = 1, X\right)$. The difference between the AIPW estimator and $\theta$ can be written as
\begin{align*}
\frac{1}{n} \sum_{i=1}^n \left\{ \frac{A_i Y_i}{\hat{\pi}} 
+ \left(1 - \frac{A_i}{\hat{\pi}}\right) \hat{m}^1(X_i)\right\}- \theta 
&= \frac{1}{n} \sum_{i=1}^n \left\{\frac{A_i Y_i}{\hat{\pi}} + \left(1 - \frac{A_i}{\hat{\pi}} \right) m^1(X_i)\right\} - \theta \\
&\quad + \frac{1}{n} \sum_{i=1}^n \left(1-\frac{A_i}{\hat{\pi}} \right) \left\{ \hat{m}^1(X_i)-m^1(X_i) \right\} \\
&= \frac{1}{n} \sum_{i=1}^n \left\{\frac{A_i Y_i}{\pi}+\left(1-\frac{A_i}{\pi} \right) m^1(X_i)\right\}-\theta \\
&\quad + \frac{1}{n} \sum_{i=1}^n \left(1 - \frac{A_i}{\hat{\pi}} \right) \left\{ \hat{m}^1(X_i)-m^1(X_i) \right\} \\
&\quad + \left( \frac{1}{\hat{\pi}} - \frac{1}{\pi} \right) \frac{1}{n} \sum_{i=1}^n A_i \left\{Y_i-m^1(X_i) \right\}.
\end{align*}

Here, the first term is asymptotically normal and $O_p(n^{-1/2})$. Further,
\begin{align*}
\frac{1}{\hat{\pi}} - \frac{1}{\pi} = O_p(n^{-1/2})
\end{align*}
and
\begin{align*}
    \frac{1}{n} \sum_{i=1}^n A_i \left\{ Y_i-m^1(X_i) \right\} = O_p(n^{-1/2})
\end{align*}
so that their product can be ignored. When the model is misspecified so that $m^1(X_i)$ is changed to be the probability limit of the estimator $\hat{m}^1(X_i)$, then this last term changes to being $O_{p}(1)$, which is no longer satisfactory. In that case, one can either make this product in the last term of the above expansion negligible using targeted learning, or work out the last term using the asymptotic distribution of $\hat{\pi}$ \citep{van2024automated}.

Finally, we study the term
\begin{equation}
\frac{1}{n} \sum_{i=1}^n \left(1-\frac{A_i}{\hat{\pi}} \right) \left\{ \hat{m}^1(X_i)-m^1(X_i)\right\} \label{eq:term}.
    \end{equation}
Suppose first that $\hat{m}^1(\cdot)$ was trained on an independent sample (denoted by $\mathcal{S}$). Then, conditional on the independent sample $\mathcal{S}$ and $\hat\pi$
\begin{align*}
    \E\left[\frac{1}{n} \sum_{i=1}^n \left(1-\frac{A_i}{\hat{\pi}} \right) \left\{ \hat{m}^1(X_i)-m^1(X_i)\right\}\ \middle| \ \mathcal{S}, \hat\pi \right]=\E\left(1-\frac{A_i}{\hat{\pi}} \ \middle| \ \hat\pi \right) \E\left[\left\{ \hat{m}^1(X_i)-m^1(X_i)\right\}\ \middle| \ \mathcal{S}\right]=0
\end{align*}
where we use that the exposures have a multivariate hypergeometric distribution conditional on $\hat{\pi}$, from which $P(A_i = 1|\hat{\pi})=\hat{\pi}$. Specifically, using Bayes' rule, $\E\left(A_i\,\middle|\, \sum_{k=1}^nA_{k}=n_1 \right)$ equals
\begin{align*}
    \E\left(A_i\,\middle|\, \sum_{k=1}^nA_{k}=n_1 \right)
    &=\frac{\E\left(\sum_{k=1}^nA_{k}=n_1\,\middle|\,A_i=1\right)\E(A_i)}{\E(\sum_{k=1}^nA_{k}=n_1)}\\
    &\quad=\frac{\E\left(\sum_{k\neq i}^nA_{k}=n_1-1\right)p}{\E(\sum_{k=1}^nA_{k}=n_1)}\\
    &\quad=\frac{\binom{n-1}{n_1-1}p^{n_1-1}(1-p)^{n-n_1}p}{\binom{n}{n_1}p^{n_1}(1-p)^{n-n_1}}\\
    &\quad=\hat{\pi}.
\end{align*}
The second equality follows from the fact that, conditional on $A_j = 1$, exactly $n_1-1$ of the remaining $n-1$ units are treated. The third equality holds because, under independent $\mathrm{Bernoulli}(p)$ assignment, the number of treated units among the remaining $n-1$ units is binomially distributed.

Further, conditional on $\mathcal{S}$ and $\hat\pi$
\begin{align*}
    &\text{Var}\left[\frac{1}{n} \sum_{i=1}^n \left(1-\frac{A_i}{\hat{\pi}} \right) \left\{ \hat{m}^1(X_i)-m^1(X_i)\right\}\ \middle| \ \mathcal{S}, \hat\pi\right]\\
&= \mathbb{E} \left( \text{Var} \left[ \frac{1}{n} \sum_{i=1}^{n} \left(1 - \frac{A_i}{\hat{\pi}}\right) \left\{ \hat{m}(X_i)-m(X_i) \right\}|X_1, \dots, X_n, \mathcal{S}, \hat{\pi} \right] \ \middle| \ \mathcal{S}, \hat\pi\right) \\
&\quad + \text{Var} \left( \mathbb{E} \left[ \frac{1}{n} \sum_{i=1}^{n} \left(1 - \frac{A_i}{\hat{\pi}}\right) \left\{ \hat{m}(X_i)-m(X_i) \right\}|X_1, \dots, X_n, \mathcal{S}, \hat{\pi} \right] \ \middle| \ \mathcal{S}, \hat\pi\right) \\
&= \mathbb{E} \left( \text{Var} \left[ \frac{1}{n} \sum_{i=1}^{n} \frac{A_i}{\hat{\pi}} \left\{ \hat{m}(X_i)-m(X_i) \right\}|X_1, \dots, X_n, \mathcal{S}, \hat{\pi} \right] \ \middle| \ \mathcal{S}, \hat\pi\right) \\
&= \mathbb{E} \left( \frac{1 - \hat{\pi}}{n \hat{\pi}} \left[ \frac{1}{n} \sum_{i=1}^{n} \left\{ \hat{m}(X_i)-m(X_i) \right\}^2-\frac{1}{n(n-1)} \sum_{i \ne j} \left\{ \hat{m}(X_i) - m(X_i) \right\} \left\{ \hat{m}(X_j)-m(X_j) \right\} \right]\ \middle| \ \mathcal{S}, \hat\pi\right),
\end{align*}
where we rely on properties of the multivariate hypergeometric distribution. This variance term converges to zero at root-$n$ rate, making \eqref{eq:term} at worst $O_p(n^{-1/2})$ by Chebychev's inequality, which is just not small enough. Indeed, the term
\begin{align*}
\frac{1}{n} \sum_{i=1}^{n} \left\{ \hat{m}(X_i)-m(X_i)\right\}^2 - \frac{1}{n(n-1)} \sum_{i,j: i \ne j} \left\{ \hat{m}(X_i)-m(X_i) \right\} \left\{ \hat{m}(X_j)-m(X_j) \right\} \tag{2}
\end{align*}
will not shrink with sample size, because it reflects a difference between a variance and covariance which is generally positive (by Cauchy-Schwarz), unless the predictions are consistent in the sense that
\begin{align*}
\frac{1}{n} \sum_{i=1}^{n} \left\{ \hat{m}(X_i)-m(X_i) \right\}^2 
\end{align*}
converges to zero. This gives little hope that \eqref{eq:term} can be ignored. 

The conclusion from the above is that the standard G-computation estimators based on canonical GLMs will be $\sqrt n$-consistent under proportional asymptotics, but that what we now use as the influence function, is in fact not the influence function since it is missing \eqref{eq:term} that we cannot approximate well.

\subsection{Leave-one-out cross-fitting based estimators}\label{appe_leave_one_out_ch3}
Leave-one-out cross-fitting based estimators have been proposed to correct the own-observation bias that arises in standard AIPW estimators when the number of covariates is large relative to the sample size \citep{zhao2024covariate,zhao2026adjusting,lee2026improving}. Before studying the bias of these estimators, we first study the bias of the standard AIPW estimators in such settings, building on the ideas of \citep{zhao2024covariate}. The standard AIPW estimator for the mean outcome under treatment can be expressed as 
\begin{align*}
  \frac{1}{n}\sum_{i=1}^{n} \left\{\frac{A_{i}}{\hat{\pi}}Y_{i}-\left(\frac{A_{i}}{\hat{\pi}}-1\right)\hat{m}^{1}( \Tilde{X}_i)
  \right\},
\end{align*}
where $\hat{m}^{1}(\Tilde{X}_i)= m(\Tilde{X}_i;\hat{\delta}_1)$ is the predicted outcome from a model of $Y$ on $\tilde{X}$ (a covariate vector with an intercept) fitted in the treated group, with estimates of regression coefficient $\hat{\delta}_1$.

For a linear model $m(\Tilde{X}_i;\hat{\delta}_1)= \Tilde{X}_i^\top \hat{\delta}_1$ fitted in the treatment group, we have
\begin{align*}
\hat{\delta}_1=\left( \Tilde{X}^\top  \Tilde{X}\right)^{-1}\sum_{j}  \Tilde{X}_j \frac{A_j}{\hat{\pi}} Y_j.
\end{align*}
Using the hat matrix $H =  \Tilde{X}\left( \Tilde{X}^\top  \Tilde{X}\right)^{-1} \Tilde{X}^\top$, the AIPW estimator for the mean outcome under treatment can be then expressed as 
\begin{align} \label{eq:tau_AIPW}
\hat{\tau}_{\text{AIPW}} = \frac{1}{n}\sum_{i=1}^{n} &\frac{A_i}{\hat{\pi}} Y_i-\frac{1}{n}\sum_{i,j}\left(\frac{A_i}{\hat{\pi}} - 1\right)H_{ij} \frac{A_j}{\hat{\pi}} Y_j,
\end{align}
where $H_{ij}=\Tilde{X}_i^\top (\Tilde{X}^\top \Tilde{X})^{-1}\Tilde{X}_j$ denotes the $(i,j)$th element of the hat matrix.

We now study the bias of the correction term in Equation \eqref{eq:tau_AIPW}, since the first term is known to be unbiased. In what follows, we consider centered covariates and omit the intercept from the hat matrix. The correction term in Equation \eqref{eq:tau_AIPW} can be decomposed as
\begin{align}\label{eq:AIPW_own_bias} 
\frac{1}{n}\sum_{i,j}\left(\frac{A_i}{\hat{\pi}}-1\right)H_{ij} \frac{A_j}{\hat{\pi}} Y_j
=\frac{1}{n}\sum_{i,j: i \ne j}\left(\frac{A_i}{\hat{\pi}}-1\right)H_{ij} \frac{A_j}{\hat{\pi}} Y_j+\frac{1}{n}\sum_{i=1}^n\left(\frac{A_i}{\hat{\pi}}-1\right)H_{ii} \frac{A_i}{\hat{\pi}} Y_i.
\end{align}
Since $\hat{\pi}$ is random, given by $\sum_{k=1}^nA_{k}/n$, we condition on it when computing the expected value of the first term in Equation \eqref{eq:AIPW_own_bias} as
 \begin{align*}
     &\E\left\{\frac{1}{n}\sum_{i,j: i \ne j} \left(\frac{A_i}{\hat{\pi}}-1\right) H_{ij} \frac{A_j}{\hat{\pi}} Y_j\,\middle|\, \sum_{k=1}^nA_{k}=n_1\right\}\\
     &\quad= \frac{1}{n}\sum_{i,j: i \ne j}\E\left\{\left(H_{ij} \frac{A_iA_j}{\hat{\pi}^2} Y_j-H_{ij} \frac{A_j}{\hat{\pi}} Y_j\right)\,\middle|\, \sum_{k=1}^nA_{k}=n_1\right\}\\
    &\quad= \frac{1}{n\hat{\pi}^2} \sum_{i,j: i \ne j}\E\left(H_{ij}Y_j^1\right)\E\left(A_iA_j\,\middle|\, \sum_{k=1}^nA_{k}=n_1 \right)-\frac{1}{n\hat{\pi}}\sum_{i,j: i \ne j}\E\left(H_{ij}Y_j^1\right)\E\left(A_j\,\middle|\, \sum_{k=1}^n A_{k}=n_1\right).
 \end{align*}
The second equality follows from consistency, the i.i.d. assumption, and random treatment assignment. In particular, treatment assignment is independent of the baseline covariates and the potential outcome, allowing the expectations involving $A$ to be separated from those involving $HY^1$.

As shown before, $\E\left(A_j\,\middle|\, \sum_{k=1}^nA_{k}=n_1 \right)=\hat{\pi}$, and similarly, by Bayes' rule
\begin{align*}
    \E\left(A_iA_j\,\middle|\, \sum_{k=1}^nA_{k}=n_1 \right) &=\frac{\E\left(\sum_{k=1}^nA_{k}=n_1\,\middle|\,A_i=A_j=1\right)\E(A_iA_j)}{\E(\sum_{k=1}^nA_{k}=n_1)}\\
    &\quad=\frac{\E\left(\sum_{k\neq i;k\neq j}^nA_{k}=n_1-2\right)p^2}{\E(\sum_{k=1}^nA_{k}=n_1)}\\
    &\quad=\frac{\binom{n-2}{n_1-2}p^{n_1-2}(1-p)^{n-n_1}p^2}{\binom{n}{n_1}p^{n_1}(1-p)^{n-n_1}}\\
    &\quad=\frac{\hat{\pi}(\hat{\pi}n-1)}{n-1}.
\end{align*}

Thus, the expected value of the first term in Equation \eqref{eq:AIPW_own_bias} can be simplified as  
\begin{align*}
   & \frac{1}{n\hat{\pi}^2} \sum_{i,j: i \ne j}\E\left(H_{ij}Y_j^1\right)\E\left(A_iA_j\,\middle|\, \sum_{k=1}^nA_{k}=n_1 \right)-\frac{1}{n\hat{\pi}}\sum_{i,j: i \ne j}\E\left(H_{ij}Y_j^1\right)\E\left(A_j\,\middle|\, \sum_{k=1}^n A_{k}=n_1\right)\\
   &\quad =\frac{\hat{\pi}n-1}{\hat{\pi}n(n-1)} \sum_{i,j: i \ne j}\E\left(H_{ij}Y_j^1\right)-\frac{1}{n}\sum_{i,j: i \ne j}\E\left(H_{ij}Y_j^1\right)\\
   &\quad =\left\{\frac{\hat{\pi}n-1}{\hat{\pi}n(n-1)}-\frac{1}{n}\right\} \sum_{i,j: i \ne j}\E\left(H_{ij}Y_j^1\right)\\
   &\quad =-\frac{1-\hat{\pi}}{\hat{\pi}n(n-1)} \sum_{i,j: i \ne j}\E\left(H_{ij}Y_j^1\right)\\
 &\quad =\frac{1-\hat{\pi}}{\hat{\pi}(n-1)}\E\left(H_{ii}Y_i^1\right)\\
  &\quad =\frac{1-\hat{\pi}}{\hat{\pi}(n-1)}\E\left\{H_{ii}\E(Y_i^1| \Tilde{X})\right\}.
\end{align*}
The fourth equality follows from the property that, for a centered projection matrix $H$, $\sum_{i \ne j} H_{ij} = -H_{ii}$. The final equality follows by the law of iterated expectations.

Furthermore, the expected value of the second term in Equation \eqref{eq:AIPW_own_bias} (conditional on $\hat{\pi}$) can be written as  
 \begin{align*}
     &\E\left\{\frac{1}{n}\sum_{i=1}^n \left(\frac{A_i}{\hat{\pi}}-1\right) H_{ii} \frac{A_i}{\hat{\pi}} Y_i\,\middle|\, \sum_{k=1}^nA_{k}=n_1\right\}\\
     &\quad= \frac{1}{n}\sum_{i=1}^n\E\left\{\left(H_{ii} \frac{A_i}{\hat{\pi}^2} Y_i-H_{ii} \frac{A_i}{\hat{\pi}} Y_i\right)\,\middle|\, \sum_{k=1}^nA_{k}=n_1\right\}\\
    &\quad= \frac{1}{n\hat{\pi}^2} \sum_{i=1}^n\E\left(H_{ii}Y_i^1\right)\E\left(A_i\,\middle|\, \sum_{k=1}^nA_{k}=n_1 \right)-\frac{1}{n\hat{\pi}}\sum_{i=1}^n\E\left(H_{ii}Y_i^1\right)\E\left(A_i\,\middle|\, \sum_{k=1}^n A_{k}=n_1\right)\\
    &\quad= \left\{\frac{1}{\hat{\pi}}-1\right\}\E\left(H_{ii}Y_i^1\right)\\
    &\quad= \frac{1-\hat{\pi}}{\hat{\pi}}\E\left\{H_{ii}\E\left(Y_i^1| \Tilde{X}\right)\right\}.
 \end{align*}
Finally, we have
\begin{align*}
    \E\left\{ \frac{1}{n}\sum_{i\neq j}\left(\frac{A_i}{\hat{\pi}}-1\right)H_{ij} \frac{A_j}{\hat{\pi}} Y_j+\frac{1}{n}\sum_{i=1}^n\left(\frac{A_i}{\hat{\pi}}-1\right)H_{ii} \frac{A_i}{\hat{\pi}} Y_i\right\}&= \left\{\frac{1-\hat{\pi}}{\hat{\pi}(n-1)}+\frac{1-\hat{\pi}}{\hat{\pi}}\right\}\E\left\{H_{ii}\E\left(Y_i^1| \Tilde{X}\right)\right\}\\
      &= \frac{(1-\hat{\pi})n}{\hat{\pi}(n-1)}\E\left\{H_{ii}\E\left(Y_i^1| \Tilde{X}\right)\right\}.
\end{align*}
Therefore, under the condition that $\E(Y_i^1| \Tilde{X})$ is bounded, the bias of $\hat{\tau}_{\text{AIPW}}$ is of order $O\!\left(\frac{n_0}{n_1}\frac{p}{n}\right)$ since $\frac{1}{n}\sum_{i=1}^n H_{ii}=p/n$. This additional bias term, of order $\frac{n_0}{n_1}\frac{p}{n}$, is commonly referred to as own-observation bias \citep{newey2018cross}.

We now examine the bias of the leave-one-out cross-fitting AIPW estimator, which coincides with the estimator proposed by \citet{zhao2024covariate}. By excluding the diagonal term from the correction, this estimator takes the form
\begin{align*}
\hat{\tau}^\dagger_{\text{HOIF}}=\frac{1}{n}\sum_{i=1}^{n} \left\{\frac{A_{i}}{\hat{\pi}}Y_{i}-\left(\frac{A_{i}}{\hat{\pi}}-1\right)
\left(\sum_{j=1}^{n} H_{ij} \frac{A_j}{\hat{\pi}}Y_j-H_{ii}\frac{A_{i}}{\hat{\pi}}Y_{i}\right)\right\},
\end{align*}
Since the correction term now retains only the off-diagonal contributions, and as established before,
\begin{align*}
   &\E\left\{\frac{1}{n}\sum_{i,j: i \ne j} \left(\frac{A_i}{\hat{\pi}}-1\right) H_{ij} \frac{A_j}{\hat{\pi}} Y_j\,\middle|\, \sum_{k=1}^nA_{k}=n_1\right\}=\frac{1-\hat{\pi}}{\hat{\pi}(n-1)}\E\left\{H_{ii}\E(Y_i^1| \Tilde{X})\right\}.
\end{align*}
Consequently, under regularity conditions, when $\E(Y_i^1| \Tilde{X})$ is bounded on the observed data, $\hat{\tau}^\dagger_{\text{HOIF}}$ has bias of order $O\!\left(\frac{n_0}{n_1}\frac{p}{n^2}\right)$. This bias is of smaller order than that of $\hat{\tau}_{\text{AIPW}}$.

Although leave-one-out cross-fitting substantially reduces own-observation bias, the remaining bias can be eliminated by centering the outcome. Consider the estimator
\begin{align*}
\hat{\tau}^\ddagger_{\text{HOIF}}=\frac{1}{n}\sum_{i=1}^{n} \left\{\frac{A_{i}}{\hat{\pi}}Y_{i}-\left(\frac{A_{i}}{\hat{\pi}}-1\right)\left(\sum_{j=1}^{n} H_{ij} \frac{A_j}{\hat{\pi}}(Y_j-C_{1})-H_{ii}\frac{A_{i}}{\hat{\pi}}(Y_i-C_{1})\right)\right\},
\end{align*}
where $C_1= \frac{\sum_{i=1}^{n}A_iH_{ii}Y_i}{\sum_{i=1}^{n}A_iH_{ii}}$. By following the same procedure above,
 \begin{align*}
      \E\left\{ \frac{1}{n}\sum_{i,j: i \ne j} \left(\frac{A_i}{\hat{\pi}}-1\right) H_{ij} \frac{A_j}{\hat{\pi}} \left(Y_j-C_1\right)\,\middle|\, \sum_{k=1}^nA_{k}=n_1\right\}
    &=\frac{1-\hat{\pi}}{\hat{\pi}(n-1)}\E\left[H_{ii}\E\left\{\left(Y_i^1-C_1\right)| \Tilde{X}\right\}\right]\\
    &=0,
 \end{align*}
This implies that the correction term in $\hat{\tau}^\ddagger_{\text{HOIF}}$ has zero expectation, meaning it will not introduce bias into the estimator. This estimator can be seen as a bias-free version of $\hat{\tau}^\dagger_{\text{HOIF}}$.

\subsection{Variance estimation for HOIF-based estimators}\label{appe_variance_estimation_HOIF_ch3}
The variance of the the HOIF-based estimator $\hat{\tau}^\dagger_{\text{HOIF}}$, corresponding to the mean outcome under treatment, can be estimated as $1/n$ times the sample variance of its empirical influence function, given by
\begin{align*}
\frac{A_{i}}{\hat{\pi}}Y_{i}-\left(\frac{A_{i}}{\hat{\pi}}-1\right)\left(\sum_{j=1}^{n} H_{ij} \frac{A_j}{\hat{\pi}}Y_j- H_{ii}\frac{A_{i}}{\hat{\pi}}Y_{i}\right)-\hat{\tau}^\dagger_{\text{HOIF}}.
\end{align*}
Similarly, for the centered version $\hat{\tau}^\ddagger_{\text{HOIF}}$, the variance estimator is obtained as $1/n$ times the sample variance of
\begin{align*}
\frac{A_{i}}{\hat{\pi}}(Y_{i}-\hat{\tau}^\ddagger_{\text{HOIF}})-\left(\frac{A_{i}}{\hat{\pi}}-1\right) \left\{\sum_{j=1}^{n} H_{ij} \frac{A_j}{\hat{\pi}}(Y_j-C_{1})- H_{ii}\frac{A_{i}}{\hat{\pi}}(Y_i-C_{1})\right\}.
\end{align*}
See also \cite{zhao2024covariate} for details.

\subsection{Sandwich variance estimators}\label{appe_sandwich_variance_ch3}
We explore the use of degree of freedom corrections to default sandwich estimators in the context of main effects linear outcome models under which $\hat{\tau}_{\text{GComp}}=\hat{\beta}$ (the estimated regression coefficient for treatment), as recommended by the FDA \citep{FDA2023} when the model does not include treatment by covariate interactions. We consider both HC1 and HC3 sandwich variance estimators, which apply degrees of freedom corrections \citep{long2000using, hayes2007using}. The HC1 estimator can be expressed as
\begin{align*}
\text{HC1} = \frac{n}{n-p-1} (Z'Z)^{-1}Z' \operatorname{diag} \left(e^2 \right) Z (Z'Z)^{-1},
\end{align*}
where $Z=(\mathbf{1},A,X)$ is the design matrix and $\operatorname{diag} \left(e^2 \right)$ represents a diagonal matrix with the squared OLS residuals
on the main diagonal. Similarly, the HC3 estimator can be expressed as 
\begin{align*}
\text{HC3} = (Z'Z)^{-1}Z' \operatorname{diag} \left\{\frac{e_i^2}{(1-h_{ii})^2} \right\}Z (Z'Z)^{-1},\ i=1\dots n
\end{align*}
where $h_{ii}$ denotes the diagonal elements of the hat matrix $Z(Z'Z)^{-1}Z'$ and $\operatorname{diag} \left\{\frac{e_i^2}{(1-h_{ii})^2}\right\}$ represents a diagonal matrix with the squared OLS residuals weighed by $1/(1-h_{ii})^2$ on the main diagonal; see \cite{hayes2007using}.

\subsection{Cross-fitting with variable selection}\label{appe_cross_fitting_variable_selection_ch3}
Because of the slow convergence of the nuisance parameter estimates when covariate-to-sample size ratio is large (as they are fitted on limited observations, without the held-out fold), sample splitting alone may increase the variability of the estimated ATE. However, incorporating variable selection may help to mitigate this increase in variance. Thus, we consider a combination of sample splitting and variable selection to address the issue of overfitting while controlling variability in ATE estimation. This can be implemented as follows:
\begin{enumerate}
\item Split the data into $K$ mutually exclusive folds ($\mathcal{I}_{k}, k=1,\dots,K$).
\item For each fold $k$:
\begin{enumerate}
    \item Apply variable selection as explained in the main paper in the folds $\mathcal{I}^c_{k}=\{1,\dots,n\}\setminus \mathcal{I}_{k}$.
    \item Construct an MLE of $\Gamma$ using the selected covariates (i.e., fit GLMs of the outcome on treatment and LASSO-selected covariates) in the folds $\mathcal{I}^c_{k}=\{1,\dots,n\}\setminus \mathcal{I}_{k}$.
    \item Compute $\hat{m}^{a}_{k,Lasso}(X)$, for $a=0,1$, based on $\widehat{\Gamma}^{\text{post-Lasso}}$ in the fold $\mathcal{I}_{k}$.
\end{enumerate}
\item The cross-fitting combined with post-LASSO G-computation estimator can be then computed as 
\begin{align*}
    \hat{\tau}_{\text{CF\_Lasso}}=\frac{1}{n}\sum_{k=1}^{K}\sum_{i=1}^{n_{k}}\left\{\hat{m}^{1}_{k,Lasso}(X_i)-\hat{m}^{0}_{k,Lasso}(X_i)\right\}.
\end{align*}
\end{enumerate}

\subsection{Best subset selection}\label{appe_best_subset_selection_ch3}
In the data analysis, we employ best subset selection to assess the performance of estimators across different subsets. Best subset selection is used to identify the most relevant subset of baseline covariates from a larger set \citep{hastie2020best}. Under a linear model, it finds the subset of $S$ covariates that produces the best fit in terms of mean squared error, solving the nonconvex problem
\begin{align*}
    \min_{\Gamma \in \mathbb{R}^{p+2}} \|Y-\Gamma'Z\|_2^2 \quad \text{subject to} \quad \|\Gamma\|_0 \leq S,
\end{align*}
where $\|Y-\Gamma'Z\|_2^2$ represents the sum of squared differences between the elements of the vector $Y$ and the linear transformation $\Gamma'Z$, and $\|\Gamma\|_0 = \sum_{i=1}^{p+2} \mathcal{I}\{\Gamma_i \neq 0\}$ is the $\ell_0$ norm of $\Gamma$ with $\mathcal{I}\{\cdot\}$ denoting the indicator function, equal to 1 if the condition is true and 0 otherwise. 

One can compare different subsets using prediction error calculated as 
\begin{align*}
    \frac{1}{n} \sum_{i=1}^{n} \frac{e_i^2}{(1-h_i)^2},
\end{align*}
where $e_i^2$ and $h_i$ represent the squared OLS residual and the diagonal element of the hat matrix, respectively. We implement best subset selection in \texttt{R} using the \texttt{bs} function from the \textit{bestsubset} package.

\section{Additional simulation and data analysis results}
\subsection{Additional simulation results}\label{appe_sim_results_ch3}
\begin{figure}[!h]
    \centering
    \includegraphics[width=1\textwidth]{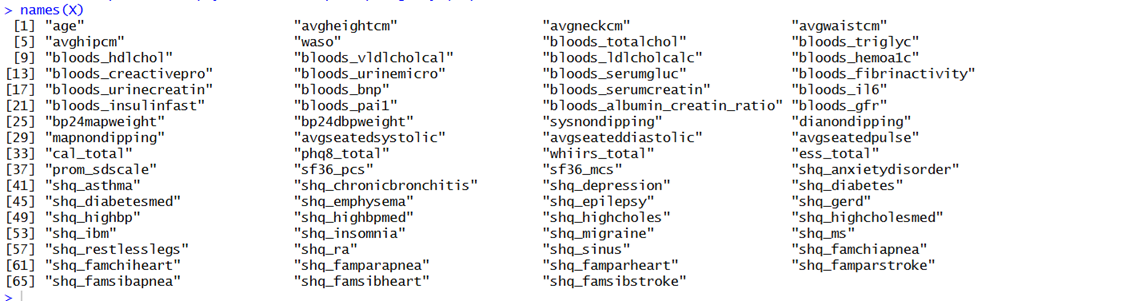}
   \caption{Covariate names in the BestAIR data set.} 
    \label{figure:appe_covariates_names_ch3}
\end{figure}

For each value of $k$ in simulation Setting 3, we include the first $k \times n$ covariates from the BestAIR data set, whose names are listed in \autoref{figure:appe_covariates_names_ch3}.

In this section, we provide additional simulation results. The sandwich variance estimators of the linear regression treatment effect coefficient work well at all values of $k$ (see \autoref{table:appe_cont_s1_WCI_HC_SE_corr_COV_ch3}). The HC1-correction gives better results than the HC3-correction when the number of covariates becomes large. These variance estimators are valid because they are not influenced by the dependence between nuisance estimation errors and the data used for ATE estimation. The small-sample correction fails to restore nominal coverage for G-computation estimators (see \autoref{table:appe_cont_s1_WCI_HC_SE_corr_COV_ch3}). This is likely because, when $k$ is large, the outcome models become overfit, leading to excessive shrinkage of the residuals and consequently underestimated variances.

\begin{table}[!h]
    \caption{Average width of 95\% confidence intervals and coverage probabilities for ATE based on sandwich variance estimators and small sample corrections, for continuous outcomes under Setting 1.}
    \begin{center}
        \resizebox{\textwidth}{!}{
            \begin{tabular}{ccccccccccccc}
                \toprule
                & \multicolumn{6}{c}{\textbf{WCI}} & \multicolumn{6}{c}{\textbf{COV (\%)}} \\
                \cmidrule(lr){2-7} \cmidrule(lr){8-13}
                & \multicolumn{2}{c}{\textbf{HC1}} & \multicolumn{2}{c}{\textbf{HC3}} & \multicolumn{2}{c}{\textbf{SE\_corr}} & \multicolumn{2}{c}{\textbf{HC1}} & \multicolumn{2}{c}{\textbf{HC3}} & \multicolumn{2}{c}{\textbf{SE\_corr}} \\
                \cmidrule(lr){2-3} \cmidrule(lr){4-5} \cmidrule(lr){6-7} \cmidrule(lr){8-9} \cmidrule(lr){10-11} \cmidrule(lr){12-13}
$k$ & $\hat{\tau}_{\text{\tiny{GComp}}}$ & $\hat{\tau}_{\text{\tiny{Lasso}}}$ & $\hat{\tau}_{\text{\tiny{GComp}}}$ & $\hat{\tau}_{\text{\tiny{Lasso}}}$ & $\hat{\tau}_{\text{\tiny{GComp}}}$ & $\hat{\tau}_{\text{\tiny{Lasso}}}$ & {$\hat{\tau}_{\text{\tiny{GComp}}}$} & {$\hat{\tau}_{\text{\tiny{Lasso}}}$} & {$\hat{\tau}_{\text{\tiny{GComp}}}$} & {$\hat{\tau}_{\text{\tiny{Lasso}}}$} & {$\hat{\tau}_{\text{\tiny{GComp}}}$} & {$\hat{\tau}_{\text{\tiny{Lasso}}}$} \\
                \midrule
                \multicolumn{13}{c}{$n$=50} \\
0.05 & 1.419 & 1.419 & 1.504 & 1.504 & 1.380 & 1.380 & 92.7 & 92.7 & 94.5 & 94.5 & 92.2 & 92.2 \\
  0.1 & 1.564 & 1.563 & 1.700 & 1.698 & 1.488 & 1.488 & 95.2 & 95.3 & 96.8 & 96.8 & 93.9 & 93.9 \\
  0.2 & 1.852 & 1.835 & 2.156 & 2.123 & 1.664 & 1.658 & 92.2 & 91.8 & 96.4 & 96.4 & 89.9 & 89.4 \\
  0.3 & 2.116 & 2.040 & 2.665 & 2.507 & 1.766 & 1.743 & 94.7 & 94.7 & 98.0 & 97.9 & 89.2 & 88.6 \\
  0.4 & 2.414 & 2.195 & 3.340 & 2.839 & 1.853 & 1.801 & 94.5 & 93.4 & 99.1 & 98.7 & 85.0 & 83.6 \\
  0.7 & 3.881 & 2.396 & 8.283 & 3.339 & 2.012 & 1.869 & 91.4 & 95.4 & 99.9 & 99.9 & 66.9 & 71.4 \\
                \midrule
                \multicolumn{13}{c}{$n$=100} \\
0.05 & 1.079 & 1.079 & 1.121 & 1.121 & 1.054 & 1.054 & 94.8 & 94.7 & 95.7 & 95.6 & 94.8 & 94.7 \\
  0.1 & 1.234 & 1.233 & 1.320 & 1.319 & 1.171 & 1.171 & 94.9 & 94.9 & 96.2 & 96.2 & 93.3 & 93.3 \\
  0.2 & 1.452 & 1.445 & 1.655 & 1.644 & 1.299 & 1.296 & 93.7 & 93.7 & 97.4 & 97.3 & 90.4 & 90.2 \\
  0.3 & 1.669 & 1.619 & 2.046 & 1.953 & 1.392 & 1.375 & 95.0 & 94.7 & 98.1 & 98.2 & 90.7 & 90.0 \\
  0.4 & 1.900 & 1.747 & 2.538 & 2.212 & 1.469 & 1.426 & 95.4 & 94.6 & 98.9 & 99.1 & 88.4 & 87.1 \\
  0.7 & 2.970 & 1.888 & 5.857 & 2.540 & 1.604 & 1.469 & 92.0 & 96.1 & 100.0 & 100.0 & 69.5 & 73.1 \\
                \midrule
                \multicolumn{13}{c}{$n$=200} \\
0.05 & 0.856 & 0.856 & 0.883 & 0.883 & 0.835 & 0.835 & 95.5 & 95.5 & 96.0 & 96.0 & 94.6 & 94.6 \\
  0.1 & 0.975 & 0.975 & 1.035 & 1.035 & 0.924 & 0.924 & 95.3 & 95.3 & 96.6 & 96.6 & 93.7 & 93.7 \\
  0.2 & 1.157 & 1.154 & 1.306 & 1.300 & 1.034 & 1.032 & 94.8 & 94.8 & 97.0 & 97.0 & 91.8 & 91.8 \\
  0.3 & 1.320 & 1.293 & 1.598 & 1.547 & 1.106 & 1.095 & 94.0 & 93.7 & 97.6 & 97.7 & 89.4 & 88.6 \\
  0.4 & 1.509 & 1.410 & 1.979 & 1.778 & 1.168 & 1.135 & 94.7 & 95.0 & 98.8 & 98.8 & 85.4 & 83.8 \\
  0.7 & 2.403 & 1.545 & 4.547 & 2.069 & 1.304 & 1.186 & 93.5 & 96.2 & 100.0 & 100.0 & 71.7 & 72.7 \\
                \bottomrule
            \end{tabular}
        }
    \end{center}
    \label{table:appe_cont_s1_WCI_HC_SE_corr_COV_ch3}
\end{table}

\clearpage
\autoref{table:appe_cont_s2_bias_SD_SE_COV_ch3} shows estimated bias, Monte Carlo standard deviation, average standard errors, and coverage probabilities for ATE obtained from Setting 2 with continuous outcomes.
\begin{table}[!h]
\caption{Estimated bias, Monte Carlo standard deviation, average standard errors, and coverage probabilities for ATE obtained from Setting 2 with continuous outcomes.}
    \begin{center}
        \resizebox{\textwidth}{!}{
            \begin{tabular}{ccccccccccccccccccc}
\toprule
$k$ & $\hat{\tau}_{\text{\tiny{Unadj}}}$ & $\hat{\tau}_{\text{\tiny{GComp}}}$ & $\hat{\tau}_{\text{\tiny{Lasso}}}$ & $\hat{\tau}_{\text{\tiny{CF}}}$ & $\hat{\tau}_{\text{\tiny{CF\_Lasso}}}$ & $\hat{\tau}^\dagger_{\text{\tiny{HOIF}}}$ & $\hat{\tau}^\ddagger_{\text{\tiny{HOIF}}}$ & $\hat{\tau}_{\text{\tiny{JASA}}}$ & $\hat{\tau}_{\text{\tiny{JASA\_cal}}}$ & $\hat{\tau}_{\text{\tiny{Unadj}}}$ & $\hat{\tau}_{\text{\tiny{GComp}}}$ & $\hat{\tau}_{\text{\tiny{Lasso}}}$ & $\hat{\tau}_{\text{\tiny{CF}}}$ & $\hat{\tau}_{\text{\tiny{CF\_Lasso}}}$ & $\hat{\tau}^\dagger_{\text{\tiny{HOIF}}}$ & $\hat{\tau}^\ddagger_{\text{\tiny{HOIF}}}$ & $\hat{\tau}_{\text{\tiny{JASA}}}$ & $\hat{\tau}_{\text{\tiny{JASA\_cal}}}$ \\
& \multicolumn{9}{c}{\textbf{Bias}} & \multicolumn{9}{c}{\textbf{SD}} \\
\cmidrule(lr){2-10} \cmidrule(lr){11-19}
\multicolumn{19}{c}{$n$=50} \\
0.05 & -0.029 & -0.006 & -0.006 & -0.007 & -0.007 & 0.013 & -0.007 & 0.027 & 0.042 & 0.602 & 0.351 & 0.351 & 0.352 & 0.352 & 0.367 & 0.361 & 0.351 & 0.369 \\
  0.1 & 0.003 & -0.009 & -0.010 & -0.010 & -0.010 & 0.016 & -0.002 & 0.015 & 0.040 & 0.667 & 0.412 & 0.413 & 0.413 & 0.413 & 0.431 & 0.422 & 0.416 & 0.423 \\
  0.2 & 0.016 & 0.011 & 0.013 & 0.011 & 0.012 & 0.030 & 0.010 & 0.021 & 0.040 & 0.685 & 0.445 & 0.443 & 0.446 & 0.444 & 0.468 & 0.459 & 0.454 & 0.460 \\
  0.3 & 0.002 & 0.016 & 0.016 & 0.016 & 0.017 & 0.030 & 0.013 & 0.031 & 0.042 & 0.658 & 0.474 & 0.459 & 0.479 & 0.455 & 0.494 & 0.481 & 0.497 & 0.495 \\
  0.4 & -0.009 & -0.025 & -0.013 & -0.022 & -0.013 & -0.006 & -0.017 & 0.017 & 0.023 & 0.631 & 0.522 & 0.478 & 0.534 & 0.476 & 0.499 & 0.490 & 0.510 & 0.495\\
  0.7 & -0.003 & 0.008 & -0.013 & 0.051 & -0.002 & 0.007 & 0.001 & 0.023 & 0.001 & 0.673 & 0.749 & 0.518 & 1.016 & 0.498 & 0.588 & 0.581 & 0.579 & 0.581 \\
\multicolumn{19}{c}{$n$=100} \\
0.05 & -0.019 & -0.004 & -0.004 & -0.005 & -0.005 & 0.004 & -0.006 & -0.009 & 0.006 & 0.457 & 0.279 & 0.279 & 0.279 & 0.279 & 0.283 & 0.280 & 0.285 & 0.290 \\
0.1 & 0.006 & -0.011 & -0.011 & -0.010 & -0.011 & -0.002 & -0.008 & 0.008 & 0.027 & 0.471 & 0.284 & 0.283 & 0.284 & 0.283 & 0.290 & 0.290 & 0.295 & 0.297 \\
  0.2 & 0.006 & -0.001 & -0.003 & -0.002 & -0.003 & 0.008 & 0.001 & 0.012 & 0.023 & 0.471 & 0.304 & 0.299 & 0.305 & 0.297 & 0.313 & 0.309 & 0.310 & 0.307 \\
  0.3 & 0.019 & 0.013 & 0.015 & 0.014 & 0.014 & 0.021 & 0.015 & 0.007 & 0.014 & 0.469 & 0.322 & 0.303 & 0.323 & 0.300 & 0.333 & 0.327 & 0.338 & 0.338 \\
  0.4 & 0.008 & -0.011 & -0.010 & -0.011 & -0.010 & -0.002 & -0.005 & -0.003 & -0.004 & 0.464 & 0.356 & 0.319 & 0.363 & 0.314 & 0.354 & 0.349 & 0.350 & 0.345\\
  0.7 & 0.023 & 0.008 & 0.008 & 0.005 & 0.010 & 0.026 & 0.022 & 0.016 & 0.004 & 0.481 & 0.513 & 0.331 & 0.622 & 0.325 & 0.422 & 0.418 & 0.401 & 0.395 \\
\multicolumn{19}{c}{$n$=200} \\
0.05 & 0.002 & 0.010 & 0.010 & 0.011 & 0.010 & 0.016 & 0.010 & 0.000 & 0.009 & 0.328 & 0.200 & 0.200 & 0.200 & 0.200 & 0.204 & 0.201 & 0.197 & 0.198 \\
  0.1 & -0.017 & -0.009 & -0.010 & -0.009 & -0.010 & -0.005 & -0.010 & 0.004 & 0.013 & 0.336 & 0.201 & 0.199 & 0.201 & 0.199 & 0.205 & 0.202 & 0.208 & 0.208 \\
  0.2 & -0.005 & 0.005 & 0.006 & 0.005 & 0.005 & 0.008 & 0.004 & 0.005 & 0.012 & 0.322 & 0.204 & 0.197 & 0.205 & 0.197 & 0.210 & 0.205 & 0.225 & 0.225 \\
  0.3 & -0.018 & -0.004 & -0.004 & -0.003 & -0.004 & -0.004 & -0.008 & 0.009 & 0.013 & 0.321 & 0.224 & 0.207 & 0.226 & 0.205 & 0.227 & 0.223 & 0.236 & 0.234 \\
  0.4 & -0.001 & -0.010 & -0.010 & -0.008 & -0.010 & -0.001 & -0.007 & 0.012 & 0.012 & 0.328 & 0.245 & 0.214 & 0.246 & 0.208 & 0.244 & 0.240 & 0.249 & 0.248 \\
  0.7 & 0.006 & 0.013 & 0.003 & 0.010 & 0.004 & 0.011 & 0.007 & -0.002 & -0.013 & 0.325 & 0.366 & 0.219 & 0.410 & 0.213 & 0.291 & 0.285 & 0.288 & 0.287 \\
\midrule
& \multicolumn{9}{c}{\textbf{SE}} & \multicolumn{9}{c}{\textbf{COV (\%)}} \\
\cmidrule(lr){2-10} \cmidrule(lr){11-19}
\multicolumn{19}{c}{$n$=50} \\
0.05 & 0.605 & 0.336 & 0.336 & 0.371 & 0.371 & 0.374 & 0.379 & 0.342 & 0.350 & 94.8 & 92.5 & 92.5 & 95.0 & 95.0 & 94.2 & 94.9 & 93.9 & 93.3 \\
  0.1 & 0.663 & 0.359 & 0.359 & 0.419 & 0.419 & 0.421 & 0.425 & 0.415 & 0.409 & 92.8 & 90.3 & 90.2 & 94.8 & 94.8 & 93.7 & 94.4 & 94.4 & 93.5 \\
  0.2 & 0.659 & 0.335 & 0.337 & 0.452 & 0.451 & 0.447 & 0.448 & 0.447 & 0.432 & 93.9 & 85.5 & 86.5 & 94.2 & 94.3 & 93.7 & 94.4 & 94.5 & 91.8 \\
  0.3 & 0.665 & 0.315 & 0.322 & 0.503 & 0.485 & 0.481 & 0.477 & 0.473 & 0.453 & 95.2 & 79.1 & 81.4 & 95.0 & 95.9 & 93.8 & 93.9 & 93.7 & 93.2 \\
  0.4 & 0.661 & 0.287 & 0.305 & 0.568 & 0.508 & 0.506 & 0.501 & 0.506 & 0.478 & 95.4 & 70.2 & 77.2 & 95.0 & 96.0 & 95.1 & 95.0 & 94.4 & 93.4 \\
  0.7 & 0.661 & 0.194 & 0.269 & 1.601 & 0.564 & 0.597 & 0.583 & 0.594 & 0.548 & 94.3 & 41.7 & 69.7 & 99.3 & 96.8 & 95.5 & 94.8 & 95.4 & 93.2 \\
\multicolumn{19}{c}{$n$=100} \\
0.05 & 0.468 & 0.261 & 0.261 & 0.281 & 0.281 & 0.281 & 0.282 & 0.281 & 0.280 & 94.9 & 92.9 & 92.9 & 94.7 & 94.7 & 94.8 & 94.8 & 93.4 & 93.3 \\
  0.1 & 0.466 & 0.254 & 0.255 & 0.291 & 0.291 & 0.291 & 0.290 & 0.290 & 0.287 & 94.3 & 92.5 & 92.6 & 95.6 & 95.8 & 94.6 & 95.1 & 94.5 & 93.9 \\
  0.2 & 0.465 & 0.238 & 0.242 & 0.314 & 0.305 & 0.308 & 0.306 & 0.309 & 0.302 & 95.1 & 87.0 & 87.5 & 95.4 & 94.9 & 95.2 & 95.4 & 94.6 & 94.4 \\
  0.3 & 0.469 & 0.224 & 0.234 & 0.348 & 0.320 & 0.331 & 0.328 & 0.327 & 0.318 & 94.6 & 81.2 & 85.2 & 96.2 & 95.5 & 94.8 & 95.6 & 94.6 & 93.3 \\
  0.4 & 0.466 & 0.206 & 0.225 & 0.389 & 0.329 & 0.350 & 0.345 & 0.349 & 0.336 & 94.6 & 73.9 & 83.0 & 96.4 & 95.9 & 93.7 & 93.8 & 94.8 & 93.2 \\
  0.7 & 0.467 & 0.142 & 0.206 & 0.883 & 0.347 & 0.414 & 0.405 & 0.413 & 0.391 & 93.9 & 39.1 & 76.8 & 98.6 & 95.8 & 94.8 & 95.1 & 95.5 & 94.8 \\
\multicolumn{19}{c}{$n$=200} \\
0.05 & 0.329 & 0.185 & 0.185 & 0.197 & 0.197 & 0.197 & 0.196 & 0.197 & 0.196 & 94.6 & 93.0 & 93.2 & 94.8 & 95.1 & 94.5 & 94.6 & 94.4 & 93.7 \\
  0.1 & 0.329 & 0.179 & 0.181 & 0.204 & 0.202 & 0.203 & 0.202 & 0.203 & 0.202 & 94.2 & 91.6 & 92.4 & 95.0 & 95.1 & 94.7 & 94.8 & 94.5 & 94.3 \\
  0.2 & 0.330 & 0.170 & 0.175 & 0.221 & 0.210 & 0.217 & 0.215 & 0.217 & 0.214 & 96.0 & 88.6 & 91.8 & 96.5 & 96.1 & 95.7 & 95.8 & 93.7 & 93.7 \\
  0.3 & 0.330 & 0.159 & 0.170 & 0.243 & 0.216 & 0.231 & 0.228 & 0.230 & 0.225 & 95.4 & 83.8 & 89.2 & 96.8 & 96.2 & 95.3 & 94.9 & 94.5 & 94.1 \\
  0.4 & 0.329 & 0.146 & 0.166 & 0.272 & 0.220 & 0.244 & 0.241 & 0.244 & 0.237 & 95.6 & 75.1 & 85.9 & 97.1 & 96.8 & 95.1 & 94.9 & 95.1 & 94.0 \\
  0.7 & 0.330 & 0.102 & 0.157 & 0.577 & 0.227 & 0.290 & 0.284 & 0.290 & 0.277 & 95.2 & 42.9 & 82.4 & 99.5 & 96.2 & 94.8 & 93.9 & 95.0 & 93.9 \\
                \bottomrule
            \end{tabular}
        }
    \end{center}
    \label{table:appe_cont_s2_bias_SD_SE_COV_ch3}
\end{table}

\begin{figure}[!h]
    \centering
    \includegraphics[width=1\textwidth]{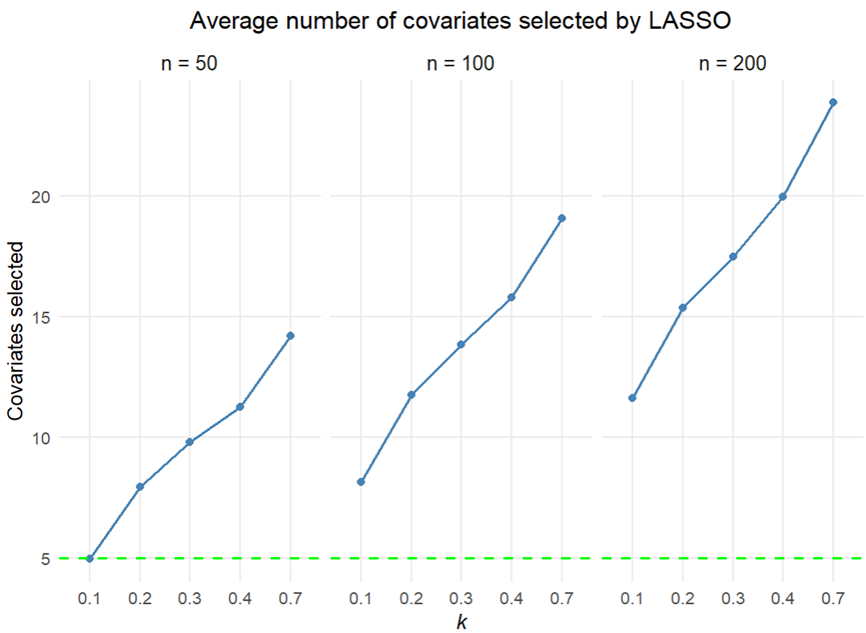}
    \caption{Average number of covariates selected by LASSO across 1000 simulations with continuous outcomes under Setting 2. The horizontal line indicates the true number of covariates with non-zero coefficients.}
    \label{figure:appe_cont_s2_p_selected_ch3}
\end{figure}

\clearpage
We found that LASSO selects more covariates than the true number of covariates with non-zero coefficients; see \autoref{figure:appe_cont_s2_p_selected_ch3}, based on the continuous outcome results from Setting 2.

\autoref{table:appe_gamma_setting3_continuous_ch3} presents the covariate coefficients ($\gamma$) considered in simulation Setting 3 for continuous outcomes.
\begin{table}[!h]
\centering
\caption{Values of $\gamma$ in Setting 3 with continuous outcomes}
        \resizebox{\textwidth}{!}{
\begin{tabular}{r r r r r r r r r r r r r r}
\hline
Index & $\gamma$ & Index & $\gamma$ & Index & $\gamma$ & Index & $\gamma$ &
Index & $\gamma$ & Index & $\gamma$ & Index & $\gamma$ \\
\hline
1  & -0.04 & 11 & -12.19 & 21 & -0.02 & 31 &  0.44 & 41 &  1.44 & 51 &  4.58 & 61 & -10.77 \\
2  & -0.01 & 12 & -1.37  & 22 &  0.49 & 32 &  0.13 & 42 & -5.04 & 52 & -3.20 & 62 &  0.59 \\
3  & -0.91 & 13 & -0.14  & 23 & -0.01 & 33 & -2.44 & 43 & -3.73 & 53 &  9.75 & 63 &  0.58 \\
4  &  0.26 & 14 &  0.01  & 24 & -0.11 & 34 & -0.15 & 44 & -0.42 & 54 &  2.90 & 64 &  0.25 \\
5  & -0.12 & 15 &  0.01  & 25 &  2.49 & 35 &  0.03 & 45 & 10.31 & 55 & -3.84 & 65 &  4.52 \\
6  & -0.01 & 16 &  0.02  & 26 & -1.86 & 36 &  0.28 & 46 &  4.03 & 56 & -12.19 & 66 &  0.38 \\
7  & 12.22 & 17 &  0.01  & 27 & -7.06 & 37 & -0.06 & 47 & -1.52 & 57 & -0.31 & 67 &  0.27 \\
8  & -1.16 & 18 &  0.00  & 28 & -2.89 & 38 &  0.09 & 48 & -1.26 & 58 &  0.81 &    &       \\
9  & -12.33 & 19 & -1.57 & 29 &  8.90 & 39 &  0.16 & 49 & -1.67 & 59 &  1.32 &    &       \\
10 & -6.52 & 20 & -0.05  & 30 & -0.02 & 40 &  1.47 & 50 &  2.05 & 60 &  1.11 &    &       \\
\hline
\end{tabular}
}
\label{table:appe_gamma_setting3_continuous_ch3}
\end{table}

\clearpage
\autoref{table:appe_bin_s2_bias_SD_SE_COV_ch3} shows estimated bias, Monte Carlo standard deviation, average standard errors, and coverage probabilities for ATE obtained from Setting 2 with binary outcomes.
\begin{table}[!h]
    \caption{Estimated bias, Monte Carlo standard deviation, average standard errors, and coverage probabilities for ATE obtained from Setting 2 with binary outcomes.}
    \begin{center}
        \resizebox{\textwidth}{!}{
\begin{tabular}{ccccccccccccccccccccccc}
\toprule
$k$ & $\E\left(Y^1\right)$ & $\E\left(Y^0\right)$ & $\hat{\tau}_{\text{\tiny{Unadj}}}$ & $\hat{\tau}_{\text{\tiny{GComp}}}$ & $\hat{\tau}_{\text{\tiny{Firth}}}$ & $\hat{\tau}_{\text{\tiny{Lasso}}}$ & $\hat{\tau}_{\text{\tiny{CF}}}$ & $\hat{\tau}_{\text{\tiny{CF\_Lasso}}}$ & $\hat{\tau}^\dagger_{\text{\tiny{HOIF}}}$ & $\hat{\tau}^\ddagger_{\text{\tiny{HOIF}}}$ & $\hat{\tau}_{\text{\tiny{JASA}}}$ & $\hat{\tau}_{\text{\tiny{JASA\_cal}}}$ & $\hat{\tau}_{\text{\tiny{Unadj}}}$ & $\hat{\tau}_{\text{\tiny{GComp}}}$ & $\hat{\tau}_{\text{\tiny{Firth}}}$ & $\hat{\tau}_{\text{\tiny{Lasso}}}$ & $\hat{\tau}_{\text{\tiny{CF}}}$ & $\hat{\tau}_{\text{\tiny{CF\_Lasso}}}$ & $\hat{\tau}^\dagger_{\text{\tiny{HOIF}}}$ & $\hat{\tau}^\ddagger_{\text{\tiny{HOIF}}}$ & $\hat{\tau}_{\text{\tiny{JASA}}}$ & $\hat{\tau}_{\text{\tiny{JASA\_cal}}}$ \\
 & & & \multicolumn{10}{c}{\textbf{Bias}} & \multicolumn{10}{c}{\textbf{SD}} \\
\cmidrule(lr){4-13} \cmidrule(lr){14-23}
\multicolumn{23}{c}{$n$=50} \\
0.05 & 0.682 & 0.547 & -0.004 & -0.003 & 0.009 & -0.004 & -0.002 & -0.002 & -0.002 & -0.003 & 0.010 & 0.010 & 0.137 & 0.126 & 0.127 & 0.128 & 0.139 & 0.140 & 0.129 & 0.130 & 0.131 & 0.131 \\
  0.1 & 0.676 & 0.545 & 0.008 & 0.008 & 0.009 & 0.009 & 0.013 & 0.011 & 0.008 & 0.008 & 0.010 & 0.009 & 0.134 & 0.119 & 0.134 & 0.123 & 0.133 & 0.136 & 0.122 & 0.124 & 0.139 & 0.141 \\
  0.2 & 0.676 & 0.546 & 0.001 & -0.004 & -0.000 & -0.002 & 0.003 & 0.002 & -0.001 & 0.001 & -0.005 & -0.005 & 0.133 & 0.125 & 0.136 & 0.128 & 0.139 & 0.139 & 0.130 & 0.129 & 0.131 & 0.136 \\
  0.3 & 0.676 & 0.546 & -0.002 & -0.009 & 0.006 & -0.004 & 0.000 & -0.002 & -0.003 & 0.001 & 0.001 & 0.001 & 0.138 & 0.134 & 0.141 & 0.136 & 0.153 & 0.152 & 0.144 & 0.138 & 0.133 & 0.133 \\
  0.4 & 0.675 & 0.545 & 0.005 & -0.005 & 0.008 & 0.003 & 0.008 & 0.007 & 0.009 & 0.006 & 0.004 & 0.004 & 0.140 & 0.137 & 0.156 & 0.136 & 0.155 & 0.148 & 0.144 & 0.141 & 0.137 & 0.141 \\
  0.7 & 0.676 & 0.546 & -0.005 & -0.040 & -0.003 & -0.009 & -0.004 & -0.002 & -0.005 & -0.004 & -0.004 & -0.002 & 0.139 & 0.118 & 0.172 & 0.135 & 0.154 & 0.151 & 0.141 & 0.140 & 0.143 & 0.143 \\
\multicolumn{23}{c}{$n$=100} \\
0.05 & 0.676 & 0.546 & 0.001 & -0.001 & -0.001 & -0.001 & -0.000 & -0.000 & -0.000 & 0.001 & -0.000 & -0.001 & 0.095 & 0.083 & 0.089 & 0.084 & 0.088 & 0.088 & 0.084 & 0.087 & 0.091 & 0.090 \\
  0.1 & 0.675 & 0.545 & 0.001 & 0.001 & 0.003 & 0.001 & 0.001 & 0.001 & 0.002 & 0.000 & 0.000 & -0.000 & 0.095 & 0.084 & 0.091 & 0.084 & 0.088 & 0.089 & 0.086 & 0.088 & 0.094 & 0.095 \\
  0.2 & 0.675 & 0.545 & -0.002 & -0.006 & 0.006 & -0.005 & -0.004 & -0.003 & -0.005 & -0.004 & 0.002 & 0.003 & 0.098 & 0.091 & 0.098 & 0.090 & 0.094 & 0.094 & 0.093 & 0.095 & 0.097 & 0.095 \\
  0.3 & 0.675 & 0.545 & 0.002 & 0.001 & 0.005 & 0.001 & 0.001 & 0.002 & 0.002 & 0.001 & -0.003 & -0.001 & 0.096 & 0.097 & 0.107 & 0.092 & 0.100 & 0.096 & 0.095 & 0.093 & 0.096 & 0.096 \\
  0.4 & 0.675 & 0.545 & 0.000 & -0.005 & 0.013 & -0.001 & -0.003 & 0.001 & -0.001 & -0.000 & -0.002 & -0.001 & 0.098 & 0.104 & 0.111 & 0.094 & 0.109 & 0.101 & 0.101 & 0.098 & 0.091 & 0.088 \\
  0.7 & 0.676 & 0.546 & -0.001 & -0.041 & 0.002 & -0.008 & -0.005 & -0.002 & -0.006 & -0.003 & 0.004 & 0.003 & 0.097 & 0.089 & 0.119 & 0.095 & 0.109 & 0.102 & 0.100 & 0.098 & 0.099 & 0.096 \\
\multicolumn{23}{c}{$n$=200} \\
0.05 & 0.675 & 0.545 & -0.001 & -0.001 & -0.000 & -0.001 & -0.002 & -0.002 & -0.001 & -0.001 & -0.000 & -0.001 & 0.071 & 0.064 & 0.062 & 0.064 & 0.065 & 0.065 & 0.065 & 0.066 & 0.063 & 0.063 \\
  0.1 & 0.675 & 0.545 & -0.003 & -0.001 & -0.001 & -0.001 & -0.001 & -0.001 & -0.001 & -0.002 & -0.002 & -0.003 & 0.071 & 0.063 & 0.063 & 0.063 & 0.065 & 0.064 & 0.064 & 0.066 & 0.066 & 0.065 \\
  0.2 & 0.675 & 0.545 & 0.002 & 0.002 & 0.002 & 0.000 & 0.000 & 0.001 & 0.001 & 0.002 & -0.003 & -0.002 & 0.070 & 0.068 & 0.066 & 0.066 & 0.070 & 0.068 & 0.067 & 0.067 & 0.065 & 0.065 \\
  0.3 & 0.676 & 0.546 & 0.001 & 0.005 & 0.009 & 0.000 & 0.001 & 0.002 & 0.000 & 0.002 & -0.001 & -0.000 & 0.069 & 0.074 & 0.073 & 0.065 & 0.075 & 0.066 & 0.068 & 0.067 & 0.066 & 0.065 \\
  0.4 & 0.676 & 0.546 & -0.005 & 0.004 & 0.013 & -0.005 & -0.007 & -0.005 & -0.005 & -0.004 & -0.000 & 0.001 & 0.069 & 0.083 & 0.084 & 0.066 & 0.076 & 0.069 & 0.068 & 0.068 & 0.066 & 0.064 \\
  0.7 & 0.676 & 0.545 & 0.002 & -0.036 & -0.005 & -0.002 & 0.001 & 0.001 & 0.002 & 0.002 & 0.000 & 0.000 & 0.069 & 0.070 & 0.089 & 0.066 & 0.076 & 0.070 & 0.070 & 0.069 & 0.069 & 0.067 \\
 \midrule
& & & \multicolumn{10}{c}{\textbf{SE}} & \multicolumn{10}{c}{\textbf{COV (\%)}} \\
\cmidrule(lr){4-13} \cmidrule(lr){14-23}
\multicolumn{23}{c}{$n$=50} \\
0.05 & 0.682 & 0.547 & 0.137 & 0.115 & 0.118 & 0.116 & 0.117 & 0.119 & 0.123 & 0.130 & 0.127 & 0.126 & 94.5 & 91.8 & 92.8 & 91.4 & 89.4 & 89.8 & 92.9 & 94.0 & 93.9 & 93.7 \\
  0.1 & 0.676 & 0.545 & 0.138 & 0.110 & 0.110 & 0.112 & 0.119 & 0.122 & 0.125 & 0.132 & 0.128 & 0.126 & 95.3 & 92.6 & 87.5 & 91.8 & 91.8 & 91.8 & 95.3 & 95.3 & 92.4 & 91.3 \\
  0.2 & 0.676 & 0.546 & 0.138 & 0.101 & 0.103 & 0.108 & 0.123 & 0.125 & 0.134 & 0.137 & 0.135 & 0.126 & 95.4 & 87.4 & 84.5 & 89.0 & 91.9 & 92.3 & 95.2 & 95.1 & 94.9 & 92.0 \\
  0.3 & 0.676 & 0.546 & 0.138 & 0.091 & 0.094 & 0.106 & 0.128 & 0.128 & 0.144 & 0.142 & 0.141 & 0.128 & 93.5 & 80.9 & 79.2 & 85.3 & 87.7 & 88.3 & 93.9 & 94.5 & 95.6 & 94.2 \\
  0.4 & 0.675 & 0.545 & 0.138 & 0.081 & 0.086 & 0.103 & 0.132 & 0.129 & 0.154 & 0.144 & 0.147 & 0.129 & 94.5 & 74.4 & 70.6 & 84.7 & 89.7 & 89.6 & 96.3 & 95.2 & 95.7 & 91.7 \\
  0.7 & 0.676 & 0.546 & 0.138 & 0.049 & 0.077 & 0.100 & 0.137 & 0.132 & 0.187 & 0.144 & 0.159 & 0.131 & 93.9 & 51.6 & 60.5 & 81.6 & 89.9 & 89.9 & 98.7 & 93.9 & 95.7 & 91.8 \\
\multicolumn{23}{c}{$n$=100} \\
0.05 & 0.676 & 0.546 & 0.097 & 0.080 & 0.080 & 0.080 & 0.083 & 0.084 & 0.085 & 0.091 & 0.087 & 0.086 & 95.5 & 95.0 & 91.0 & 94.9 & 94.1 & 94.2 & 95.9 & 96.2 & 93.5 & 93.2 \\
  0.1 & 0.675 & 0.545 & 0.097 & 0.078 & 0.078 & 0.079 & 0.086 & 0.087 & 0.088 & 0.093 & 0.090 & 0.087 & 95.3 & 93.0 & 90.9 & 93.3 & 94.8 & 94.7 & 95.7 & 96.0 & 94.1 & 93.0 \\
  0.2 & 0.675 & 0.545 & 0.097 & 0.072 & 0.073 & 0.076 & 0.091 & 0.090 & 0.095 & 0.097 & 0.095 & 0.090 & 94.9 & 86.3 & 84.0 & 88.8 & 93.7 & 93.9 & 95.4 & 94.8 & 94.4 & 93.2 \\
  0.3 & 0.675 & 0.545 & 0.097 & 0.064 & 0.066 & 0.073 & 0.097 & 0.092 & 0.101 & 0.099 & 0.100 & 0.092 & 94.5 & 80.4 & 77.5 & 87.2 & 94.1 & 93.5 & 95.9 & 96.5 & 95.8 & 93.8 \\
  0.4 & 0.675 & 0.545 & 0.097 & 0.056 & 0.060 & 0.072 & 0.101 & 0.093 & 0.108 & 0.102 & 0.104 & 0.093 & 93.7 & 69.9 & 69.6 & 84.2 & 91.7 & 92.3 & 95.8 & 94.8 & 97.9 & 96.1 \\
  0.7 & 0.676 & 0.546 & 0.097 & 0.029 & 0.053 & 0.068 & 0.104 & 0.096 & 0.131 & 0.102 & 0.112 & 0.094 & 94.7 & 38.0 & 60.3 & 82.2 & 93.4 & 93.4 & 98.7 & 96.3 & 97.9 & 94.3 \\
\multicolumn{23}{c}{$n$=200} \\
0.05 & 0.675 & 0.545 & 0.069 & 0.057 & 0.057 & 0.057 & 0.060 & 0.060 & 0.061 & 0.065 & 0.062 & 0.061 & 93.0 & 91.9 & 92.5 & 92.2 & 93.4 & 93.0 & 92.8 & 93.9 & 94.3 & 94.1 \\
  0.1 & 0.675 & 0.545 & 0.069 & 0.055 & 0.055 & 0.056 & 0.062 & 0.062 & 0.063 & 0.066 & 0.064 & 0.062 & 94.1 & 91.2 & 91.0 & 91.9 & 93.3 & 93.4 & 93.7 & 94.7 & 93.7 & 93.9 \\
  0.2 & 0.675 & 0.545 & 0.068 & 0.051 & 0.051 & 0.053 & 0.067 & 0.064 & 0.067 & 0.068 & 0.068 & 0.064 & 94.3 & 85.3 & 87.1 & 89.1 & 93.2 & 94.0 & 94.8 & 95.4 & 95.8 & 94.2 \\
  0.3 & 0.676 & 0.546 & 0.068 & 0.046 & 0.047 & 0.052 & 0.071 & 0.065 & 0.071 & 0.070 & 0.071 & 0.065 & 94.9 & 76.1 & 79.1 & 89.7 & 94.1 & 94.4 & 95.9 & 96.0 & 96.5 & 95.4 \\
  0.4 & 0.676 & 0.546 & 0.069 & 0.038 & 0.042 & 0.051 & 0.075 & 0.066 & 0.076 & 0.072 & 0.074 & 0.066 & 94.4 & 62.1 & 65.7 & 86.8 & 94.1 & 93.2 & 96.7 & 95.0 & 98.1 & 95.4 \\
  0.7 & 0.676 & 0.545 & 0.068 & 0.017 & 0.037 & 0.048 & 0.076 & 0.068 & 0.093 & 0.072 & 0.080 & 0.067 & 94.7 & 28.7 & 58.6 & 83.2 & 94.2 & 93.8 & 99.1 & 95.7 & 97.4 & 94.6 \\
                \bottomrule
            \end{tabular}
        }
    \end{center}
    \label{table:appe_bin_s2_bias_SD_SE_COV_ch3}
\end{table}

\clearpage
\autoref{table:appe_gamma_setting3_binary_ch3} presents the covariate coefficients ($\gamma$) considered in simulation Setting 3 for binary outcomes.
\begin{table}[!h]
\centering
\caption{Covariate coefficients $\gamma$ considered in simulation Setting 3 with binary outcomes.}
        \resizebox{\textwidth}{!}{
\begin{tabular}{r r r r r r r r r r r r r r}
\hline
Index & $\gamma$ & Index & $\gamma$ & Index & $\gamma$ & Index & $\gamma$ &
Index & $\gamma$ & Index & $\gamma$ & Index & $\gamma$ \\
\hline
1  & -0.06 & 11 &  0.00 & 21 &  0.03 & 31 & -0.02 & 41 &  0.69 & 51 &  3.39 & 61 & -0.49 \\
2  &  0.02 & 12 & -0.20 & 22 &  0.24 & 32 &  0.03 & 42 & -1.73 & 52 & -1.19 & 62 & -0.55 \\
3  & -0.19 & 13 &  0.01 & 23 &  0.00 & 33 &  0.45 & 43 & -0.82 & 53 &  1.16 & 63 & -1.36 \\
4  &  0.01 & 14 &  0.01 & 24 & -0.04 & 34 &  0.05 & 44 &  0.54 & 54 &  1.05 & 64 &  1.97 \\
5  & -0.02 & 15 &  0.00 & 25 &  0.66 & 35 &  0.03 & 45 &  1.21 & 55 & -0.26 & 65 &  0.22 \\
6  &  0.00 & 16 &  0.00 & 26 & -0.27 & 36 &  0.02 & 46 &  2.15 & 56 &  0.03 & 66 & -0.08 \\
7  &  0.00 & 17 &  0.00 & 27 & -0.26 & 37 &  0.08 & 47 & -0.36 & 57 &  0.89 & 67 & -0.36 \\
8  &  0.00 & 18 &  0.00 & 28 & -0.76 & 38 &  0.01 & 48 & -0.81 & 58 & -0.41 &    &       \\
9  &  0.00 & 19 & -2.32 & 29 &  1.74 & 39 &  0.01 & 49 &  2.47 & 59 &  0.08 &    &       \\
10 & -0.01 & 20 & -0.03 & 30 &  0.02 & 40 &  0.35 & 50 & -0.98 & 60 &  0.32 &    &       \\
\hline
\end{tabular}
}
\label{table:appe_gamma_setting3_binary_ch3}
\end{table}

The average width of confidence intervals and coverage probabilities for ATE based on small sample corrections, for binary outcomes under Setting 1 are summarized here (see \autoref{table:appe_bin_s1_WCI_SE_corr_COV_ch3}).
\begin{table}[!h]
    \caption{Average width of 95\% confidence intervals and coverage probabilities for ATE based on variance estimators that use small sample corrections, for binary outcomes under Setting 1.}
    \begin{center}
        \resizebox{\textwidth}{!}{
        \begin{tabular}{ccccccccccccc}
            \toprule
            & \multicolumn{6}{c}{\textbf{WCI}} & \multicolumn{6}{c}{\textbf{COV (\%)}} \\
            \cmidrule(lr){2-7} \cmidrule(lr){8-13}
            & \multicolumn{2}{c}{$n$=50} & \multicolumn{2}{c}{$n$=100} & \multicolumn{2}{c}{$n$=200} & \multicolumn{2}{c}{$n$=50} & \multicolumn{2}{c}{$n$=100} & \multicolumn{2}{c}{$n$=200} \\
            \cmidrule(lr){2-3} \cmidrule(lr){4-5} \cmidrule(lr){6-7} \cmidrule(lr){8-9} \cmidrule(lr){10-11} \cmidrule(lr){12-13}
            $k$ & {$\hat{\tau}_{\text{\tiny{GComp}}}$} & {$\hat{\tau}_{\text{\tiny{Lasso}}}$} & {$\hat{\tau}_{\text{\tiny{GComp}}}$} & {$\hat{\tau}_{\text{\tiny{Lasso}}}$} & {$\hat{\tau}_{\text{\tiny{GComp}}}$} & {$\hat{\tau}_{\text{\tiny{Lasso}}}$} & {$\hat{\tau}_{\text{\tiny{GComp}}}$} & {$\hat{\tau}_{\text{\tiny{Lasso}}}$} & {$\hat{\tau}_{\text{\tiny{GComp}}}$} & {$\hat{\tau}_{\text{\tiny{Lasso}}}$} & {$\hat{\tau}_{\text{\tiny{GComp}}}$} & {$\hat{\tau}_{\text{\tiny{Lasso}}}$} \\
            \midrule
0.05 & 0.472 & 0.471 & 0.324 & 0.324 & 0.224 & 0.224 & 95.7 & 95.6 & 93.4 & 93.0 & 94.7 & 94.6 \\
0.1  & 0.458 & 0.458 & 0.316 & 0.315 & 0.217 & 0.216 & 90.8 & 90.1 & 93.9 & 93.6 & 92.0 & 91.7 \\
0.2  & 0.439 & 0.439 & 0.299 & 0.298 & 0.206 & 0.203 & 93.1 & 91.3 & 90.2 & 89.5 & 89.5 & 88.8 \\
0.3  & 0.418 & 0.428 & 0.281 & 0.288 & 0.189 & 0.194 & 89.8 & 88.5 & 86.4 & 87.1 & 81.8 & 86.9 \\
0.4  & 0.391 & 0.421 & 0.253 & 0.282 & 0.162 & 0.188 & 85.4 & 86.8 & 79.7 & 84.7 & 71.3 & 80.9 \\
0.7  & 0.346 & 0.410 & 0.184 & 0.273 & 0.097 & 0.183 & 84.0 & 83.5 & 63.6 & 78.8 & 43.4 & 78.3 \\
            \bottomrule
        \end{tabular}
        }
    \end{center}
    \label{table:appe_bin_s1_WCI_SE_corr_COV_ch3}
\end{table}

\clearpage
\begin{table}[!h]
    \caption{Estimated power (\%) of the test of $H_0\!: \tau = 0$ at the 5\% significance level, obtained from Setting 1 with continuous outcomes.}
    \begin{center}
        \resizebox{\textwidth}{!}{
\begin{tabular}{cccccccccc}
\toprule
$k$ & $\hat{\tau}_{\text{\tiny{Unadj}}}$ & $\hat{\tau}_{\text{\tiny{GComp}}}$ & $\hat{\tau}_{\text{\tiny{Lasso}}}$ & $\hat{\tau}_{\text{\tiny{CF}}}$ & $\hat{\tau}_{\text{\tiny{CF\_Lasso}}}$ & $\hat{\tau}^\dagger_{\text{\tiny{HOIF}}}$
& $\hat{\tau}^\ddagger_{\text{\tiny{HOIF}}}$ & $\hat{\tau}_{\text{\tiny{JASA}}}$ & $\hat{\tau}_{\text{\tiny{JASA\_cal}}}$ \\ 
\cmidrule(lr){2-10}

\multicolumn{10}{c}{$n$=50} \\
0.05 & 39.6 & 80.3 & 80.3 & 74.6 & 74.6 & 74.9 & 73.9 & 84.1 & 82.3 \\ 
  0.1 & 32.8 & 77.2 & 77.2 & 69.4 & 69.4 & 69.8 & 67.7 & 69.8 & 72.8 \\ 
  0.2 & 26.7 & 68.7 & 68.1 & 49.0 & 48.2 & 51.2 & 49.9 & 54.7 & 58.1 \\ 
  0.3 & 24.2 & 68.5 & 67.7 & 40.0 & 37.8 & 44.6 & 44.1 & 46.2 & 51.2 \\ 
  0.4 & 25.7 & 69.4 & 67.8 & 29.3 & 29.3 & 39.7 & 38.1 & 38.9 & 42.5 \\ 
  0.7 & 22.0 & 75.1 & 67.2 & 2.3 & 18.5 & 25.3 & 27.2 & 26.1 & 29.1 \\ 

\multicolumn{10}{c}{$n$=100} \\
0.05 & 58.6 & 96.1 & 96.1 & 94.4 & 94.4 & 94.5 & 94.7 & 93.8 & 94.2 \\ 
  0.1 & 49.8 & 93.6 & 93.6 & 89.5 & 89.5 & 89.5 & 88.3 & 86.8 & 87.3 \\ 
  0.2 & 42.0 & 87.3 & 87.2 & 72.7 & 72.9 & 76.0 & 75.9 & 75.6 & 78.0 \\ 
  0.3 & 37.8 & 84.1 & 84.0 & 57.3 & 56.2 & 65.1 & 64.0 & 65.2 & 67.2 \\ 
  0.4 & 34.8 & 81.9 & 81.2 & 43.5 & 44.7 & 53.9 & 55.0 & 55.1 & 57.4 \\ 
  0.7 & 27.2 & 81.6 & 74.1 & 4.1 & 26.0 & 35.0 & 37.1 & 38.4 & 40.7 \\ 

\multicolumn{10}{c}{$n$=200} \\
0.05 & 79.5 & 100.0 & 100.0 & 100.0 & 100.0 & 100.0 & 100.0 & 99.6 & 99.7 \\ 
  0.1 & 65.7 & 99.1 & 99.1 & 97.5 & 97.5 & 97.4 & 97.7 & 97.8 & 97.9 \\ 
  0.2 & 59.6 & 96.2 & 96.3 & 90.6 & 90.6 & 90.7 & 90.7 & 92.0 & 93.0 \\ 
  0.3 & 53.5 & 93.2 & 93.5 & 80.1 & 79.9 & 83.4 & 83.8 & 84.5 & 85.6 \\ 
  0.4 & 46.5 & 91.4 & 90.7 & 63.6 & 64.0 & 72.7 & 73.1 & 76.0 & 77.0 \\ 
  0.7 & 39.9 & 87.2 & 85.0 & 8.6 & 41.8 & 50.9 & 51.1 & 50.8 & 55.3 \\
\bottomrule
\end{tabular}
        }
    \end{center}
    \label{table:appe_cont_s1_power_ch3}
\end{table}

\clearpage
\begin{table}[!h]
    \caption{Estimated power (\%) of the test of $H_0\!: \tau = 0$ at the 5\% significance level, obtained from Setting 1 with binary outcomes.}
    \begin{center}
        \resizebox{\textwidth}{!}{
\begin{tabular}{lcccccccccc}
  \hline
$k$ & $\hat{\tau}_{\text{\tiny{Unadj}}}$ & $\hat{\tau}_{\text{\tiny{GComp}}}$ & $\hat{\tau}_{\text{\tiny{Lasso}}}$ & $\hat{\tau}_{\text{\tiny{CF}}}$ & $\hat{\tau}_{\text{\tiny{CF\_Lasso}}}$ & $\hat{\tau}^\dagger_{\text{\tiny{HOIF}}}$ & $\hat{\tau}^\ddagger_{\text{\tiny{HOIF}}}$ & $\hat{\tau}_{\text{\tiny{Firth}}}$ & $\hat{\tau}_{\text{\tiny{JASA}}}$ & $\hat{\tau}_{\text{\tiny{JASA\_cal}}}$ \\ 
\cmidrule(lr){2-11}

\multicolumn{11}{c}{$n$=50} \\
0.05 & 16.6 & 22.5 & 22.4 & 24.7 & 23.8 & 18.5 & 17.7 & 25.2 & 23.0 & 22.9 \\ 
  0.1 & 19.4 & 25.7 & 25.5 & 26.6 & 26.2 & 20.2 & 19.2 & 29.1 & 21.6 & 23.2 \\ 
  0.2 & 15.0 & 31.1 & 27.9 & 24.5 & 23.2 & 16.1 & 17.5 & 30.8 & 13.3 & 17.5 \\ 
  0.3 & 14.6 & 35.0 & 29.5 & 22.3 & 19.9 & 14.0 & 16.4 & 40.8 & 14.6 & 19.3 \\ 
  0.4 & 15.0 & 39.0 & 30.9 & 20.7 & 20.2 & 11.3 & 16.6 & 46.4 & 13.2 & 18.6 \\ 
  0.7 & 13.1 & 49.3 & 28.1 & 18.7 & 18.1 & 4.6 & 13.4 & 44.0 & 9.4 & 17.0 \\ 

\multicolumn{11}{c}{$n$=100} \\
0.05 & 25.4 & 36.4 & 36.1 & 34.1 & 34.1 & 31.7 & 31.1 & 37.7 & 31.9 & 32.2 \\ 
  0.1 & 24.0 & 39.3 & 38.7 & 33.9 & 32.9 & 31.5 & 30.6 & 42.9 & 30.8 & 33.3 \\ 
  0.2 & 23.9 & 43.7 & 41.5 & 27.6 & 26.7 & 25.2 & 27.0 & 47.3 & 25.5 & 30.2 \\ 
  0.3 & 24.4 & 48.9 & 43.4 & 26.8 & 25.4 & 21.3 & 26.5 & 53.2 & 20.6 & 25.7 \\ 
  0.4 & 24.2 & 57.0 & 42.0 & 23.0 & 25.5 & 17.6 & 24.7 & 55.9 & 16.6 & 22.0 \\ 
  0.7 & 20.6 & 66.3 & 43.1 & 21.0 & 22.4 & 7.5 & 21.5 & 55.1 & 15.5 & 24.2 \\ 

\multicolumn{11}{c}{$n$=200} \\
0.05 & 45.8 & 62.5 & 62.5 & 57.3 & 57.2 & 57.5 & 57.6 & 60.6 & 55.0 & 55.6 \\ 
  0.1 & 39.7 & 59.5 & 59.5 & 50.5 & 51.1 & 49.0 & 50.6 & 57.9 & 48.0 & 49.0 \\ 
  0.2 & 35.5 & 62.2 & 59.8 & 40.5 & 39.5 & 39.2 & 42.1 & 65.9 & 41.2 & 45.6 \\ 
  0.3 & 35.9 & 68.7 & 61.4 & 37.1 & 38.6 & 35.2 & 41.1 & 71.2 & 38.1 & 42.4 \\ 
  0.4 & 32.7 & 75.9 & 58.6 & 32.6 & 35.2 & 28.6 & 37.6 & 74.7 & 32.1 & 37.9 \\ 
  0.7 & 33.8 & 82.6 & 58.2 & 32.7 & 33.0 & 15.8 & 35.7 & 68.1 & 25.9 & 35.1 \\ 
\bottomrule
\end{tabular}
        }
    \end{center}
    \label{table:appe_bin_s1_power_ch3}
\end{table}

\clearpage
\subsection{Additional data analyses results}\label{appe_data_analysis_results_ch3}
\begin{figure}[!h]
    \centering
    \includegraphics[width=1\textwidth]{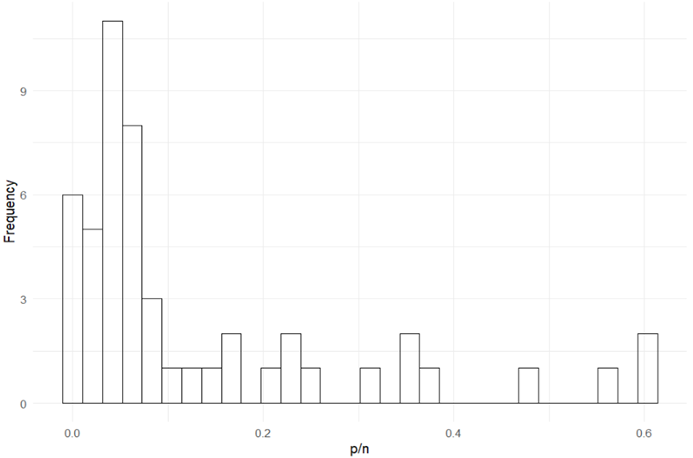}
   \caption{Distribution of the ratio of the number of covariates to the sample size across 50 RCTs, as reported by \cite{shao2026benchmarking}.} 
    \label{figure:appe_k_Bingkai_ch3}
\end{figure}

\clearpage
\bibliographystyle{plainnat}
\bibliography{bibilio}

\end{document}